\documentclass[1p,authoryear]{elsarticle}

\usepackage{amssymb}
\usepackage{natbib}
\usepackage[latin1]{inputenc}
\usepackage{amscd}
\usepackage{amsmath}
\usepackage{amsfonts}
\usepackage{bm}
\usepackage{bbm}
\usepackage{latexsym}
\usepackage{mathrsfs}
\usepackage{amsthm}
\usepackage{algorithm}
\usepackage{algpseudocode}
\usepackage{algpascal}
\usepackage{mathrsfs}
\usepackage{booktabs}
\usepackage{hyperref}
\hypersetup{linktocpage,
		colorlinks = true,
		linkcolor = blue,
		urlcolor  = blue,
		citecolor = red,
		anchorcolor = green,
}
\usepackage{rotating}
\usepackage{mathtools}
\usepackage{multirow}
\usepackage{epsfig}
\usepackage{epstopdf}

\biboptions{round}

\newcommand\bbone{\ensuremath{\mathbbm{1}}}

\def\bx{\mathbf{x}}

\def\by{\mathbf{y}} 

\def\bbeta{\mbox{\boldmath $\beta$}}

\def\bmu{\mbox{\boldmath $\mu$}}

\def\bSigma{\mathbf{\Sigma}}
\def\bXi{\mathbf{\Xi}}

\def\sE{\mathsf{E}}
\def\sS{\mathsf{S}}
\def\sP{\mathsf{P}}

\def\trasp{\mathsf{T}}

\begin{document}

\begin{frontmatter}

\title{Unified Bayesian Conditional Autoregressive Risk Measures using the Skew Exponential Power Distribution}


\author[bottone]{Marco Bottone \corref{cor1}}
\author[bernardi]{Mauro Bernardi}
\author[petrella]{Lea Petrella}

\cortext[cor1]{Corresponding author. E-mail address: marco.bottone@bancaditalia.it. The views expressed are not necessarily shared by the Bank of Italy.}
\address[bottone]{Bank of Italy, Directorate General for Economics, Statistics and Research, Italy.}
\address[bernardi]{Department of Statistical Sciences, University of Padua, Padua, Italy.}
\address[petrella]{Department of Methods and Models for Economics, Territory and Finance\\ Sapienza University of Rome, Rome, Italy.}

\begin{abstract}
Conditional Autoregressive Value--at--Risk and Conditional Autoregressive Expectile have become two popular approaches for direct measurement of market risk. Since their introduction
several improvements both in the Bayesian and in the classical framework have been proposed to better account for  asymmetry and local non--linearity. 
Here we propose a unified Bayesian Conditional Autoregressive Risk Measures approach by using the Skew Exponential Power distribution. 
Further, we extend the proposed models using a semiparametric P--spline approximation answering for a flexible way to consider the presence of non-linearity. To make the statistical inference we adapt the MCMC algorithm proposed in \cite{bernardi_bottone_petrella.2018} to our case.
The effectiveness of the whole approach is demonstrated using real data on daily return of five stock market indices.
\begin{keyword}
Bayesian quantile regression, Skew Exponential Power, risk measure, adaptive--MCMC, CAViaR model, CARE model
\end{keyword}

\end{abstract}

\end{frontmatter}

\section{Introduction}
\label{sec:intro}
%
\noindent After the recent financial crisis an accurate risk measurement is a primary need for financial institutions and investors. Within the instruments for market risk measurement, Value--at--Risk (VaR) (\cite{jorion.2007}) and Expected Shortfall (ES) (\cite{artzner_etal.1999}) are certainly the most popular and used approaches.
VaR answers the question on what is the maximum potential loss that will be exceeded with a certain probability in the next days. It can be simply understood as a specific (say $\tau$) conditional quantile of the portfolio returns given the current information, i.e., $\mathbb{P}\left(Y_t<-{\rm VaR}_t\mid\mathcal{F}_t\right)=\tau$, where $Y_t$ and $\mathcal{F}_t$ denote the return of a portfolio and the information set available at time $t$, respectively, while $\tau\mathcal{2}\left(0,1\right)$ denotes the quantile confidence level associated with the VaR. Even though it is widely used among financial institutions VaR has been criticized because of the absence of the sub--additivity property, namely, it does not guarantee that a diversified portfolio is less risky than a concentrated one. In addition VaR gives no information regarding possible exceedances beyond the quantile which may be quite important in evaluating risks. \cite{artzner_etal.1999} first recognized the lack of coherency of the VaR and proposed the ES as an alternative coherent risk measure which gives more information about the returns' distribution in the tails. In particular the ES is defined as the conditional expected loss given that the loss exceed the VaR, i.e., $\mathbb{E}\left(Y_t \mid y_t <-{\rm VaR}\right)$. Despite ES is a coherent risk measure, it is in general more difficult to calculate and to backtest it, i.e., to verify how accurately the strategy or method would have predicted actual results. Its modeling and measuring is still an ongoing problem without reaching a consensus. Moreover, it is not as simple to interpret as the VaR, and, for these reasons there is not a prevailing risk measure between VaR and ES.

\indent 
In financial time series it is usually the case that the distribution of the returns typically
changes over time, for this reason the recent literature focuses on modeling VaR and the ES by considering conditional autoregressive models. A recent development in the
VaR literature is the Conditional Autoregressive Value at Risk (CAViaR) class of models (see \cite{engle_manganelli.2004}). In particular it specifies the evolution of the quantile over time using an autoregressive approach, and estimates the parameters using a quantiles regression framework. This approach has strong appeal in that it focuses the tail of return distribution directly and does not rely on any distributional assumption. 

\indent In order to impose a dynamic evolution over ES risk measure, \cite{taylor_jw.2008} elaborates the Conditional Autoregressive Expectile (CARE) class of models which delivers estimates for both VaR and ES by considering the theory of expectiles. In practice the expectile is used as the basis for estimating ES, obtained applying the equation \eqref{eq:care2es} of Section \ref{sec:CARM_models}. Moreover, the one--to--one mapping from expectiles to quantiles can be used to obtain also an estimate of the VaR by following \cite{efron.1991} who showed that the estimator of the $\tau$--th quantile is the corresponding $\nu$--th expectile having the proportion of observations below equal to $\tau \%$.

\indent The inferential approach proposed in the literature to estimate the parameters of the two class of models consider both the frequentist and the Bayesian tools. In the CaViaR framework the former approach is based on quantile regression methods (\cite{koenker.2005}, \cite{engle_manganelli.2004}) consisting in minimizing the loss function introduced by \cite{koenker_basset.1978}. The Bayesian approach instead relies on the Asymmetric Laplace (AL) distribution tool; see for example \cite{yu_moyeed.2001}, \cite{kottas_gelfand.2001}, \cite{kottas_krnjajic.2009}, \cite{sriram_etal.2013} and \cite{bernardi_etal.2015}.
From a CARE model point of view the frequentist approach relies on the Asymmetric Least Squares (ALS) as in \cite{newey_powell.1987} while the Bayesian paradigm consider the Asymmetric Gaussian (AG) distribution assumption (see, e.g., \cite{gerlach_wang.2015}, \cite{gerlach_chen.2014},  \cite{wichitaksorn_etal.2014}, \cite{gerlach_etal.2016}, \cite{gerlach_etal.2016b}).

\indent With this paper we introduce three main innovations on the existing literature on CaViaR and CARE models. First, we develop a unified Bayesian Conditional Autoregressive Risk Measure (B--CARM) model which encompasses both the CAViaR and the CARE ones as particular cases by using the Skew Exponential Power (SEP) as working likelihood. Using the properties and the parametrization of the SEP presented in \cite{kobayashi.2015}, we show how to estimate the CAViaR and the CARE class of Conditional Autoregressive Risk models by varying one of the parameter of the SEP distribution. A parallel idea is developed in \cite{kobayashi.2015} 
where quantiles and expectiles are calculated for a stochastic volatility model when the SEP distribution is assumed. Differently from \cite{kobayashi.2015} our paper uses the properties of the SEP distribution to estimate of Conditional Autoregressive models in the risk measures framework.

\indent The second new issue of the paper consists in proposing a new nonlinear and semiparametric model specification (the BNL--CARM) of B--CARM one by using the penalized B--Splines approximation for the News Impact Curve (NIC)  (see, \cite{deboor.2001}, \cite{eilers_marx.1996},    \cite{lang_brezger.2004}) to estimate the relation between the quantile/expectile and the exogenous variables. The need for a model that allows for nonlinearity without assuming any particular shape restrictions is of clear interest in financial literature. Most of the existing literature on CaViaR and CARE imposes a priori the nonlinearity relation between the observed variables and the current quantile or expectile level 
(see i.e. \cite{engle_manganelli.2004}, \cite{gerlach_etal.2011} and \cite{gerlach_etal.2012}, \cite{chen_etal.2009} and \cite{chen_etal.2012c}, \cite{gerlach_chen.2014}). 

\indent Finally, a new Adaptive--Independent Metropolis--Hastings (AIMH) algorithm is implemented to estimate efficiently the model parameters both in the parametric and semiparametric versions. In particular modify and readapt the method proposed in \cite{bernardi_bottone_petrella.2018} to our case. The algorithm is in line with the Adaptive MCMC methods proposed in \cite{liang_etal.2010} that allows the proposal distribution to be updated at each iteration to tailor its shape to that of the target one. These methods do not require the prior specification of the proposal parameters and their theoretical properties are now well understood, see, e.g., \cite{andrieu_thoms.2008} and \cite{liang_etal.2010}, \cite{atchade_rosenthal.2005}, \cite{atchade_moulines.2011}.\newline
\indent From an applied point of view we estimate the BNL--CARM and the B--CARM models for five stock market indices comparing their performances. In particular we conduct a backtesting procedure showing that, even without imposing any restrictive assumption on the relations among variables, the proposed BNL--CARM performances are in line with those of the competitors. By that we can consider the BNL--CARM as a valid and more general alternative to the existing models. \newline
\indent The remaining of the paper is organized as follows. Section \ref{sec:CARM_models} briefly reviews conditional autoregressive risk models; Section \ref{sec:B_CARM_models} introduces the SEP as working likelihood for the Bayesian Conditional Autoregressive Risk Model with the CAViaR and the CARE models being special cases, while Section \ref{sec:BNL_CARM_model} proposes a nonlinear and semiparametric extension. Section \ref{sec:Bay_methods} shows and discusses the implemented Bayesian methododology; 
Section \ref{sec:empirical_applications} shows results from real datasets and Section \ref{sec:Conclusion} provides concluding remarks.
%
\section{CAViaR and CARE models}
\label{sec:CARM_models}
%
\noindent 
The most important VaR and ES dynamic models proposed in literature are the class of conditional autoregressive risk measure models known as the CAViaR and the CARE models introduced by \cite{engle_manganelli.2004} and \cite{taylor_jw.2008}, respectively. 
The CAViaR class of models attempt to compute the $\tau$--th level VaR by estimating the $\tau$--th level quantile of the portfolio returns through a conditional autoregressive equations structure. More specifically, let $y_t$ be the return at time $t$, the CAViaR model has the following form:
\begin{align}
y_t & = q_{t,\tau}(\theta) + \varepsilon_t \\
q_{t,\tau}(\theta) & =\omega + \gamma q_{t-1,\tau}(\theta) + \ell\left(\bbeta,y_{t-1}\right),
\label{eq:caviar_model}
\end{align}
where $q_{t,\tau}(\theta)$ is the $\tau$ level quantile of $y_{t}$ conditional upon the information set up to time $t-1$, $F_{t-1}$. It is defined as the value that minimizes the function $\mathbb{E}\left[\left\vert\tau-\bbone_{\left(-\infty,q_{t,\tau}(\theta)\right)}\left(y_t\right)\right\vert\left\vert y_t - q_{t,\tau}(\theta) \right\vert\right]$, where $\bbone_A$ is the indicator function of the set $A$, $\theta=\left(\omega, \gamma, \bbeta \right)$ are the parameters of the model and $l\left(\cdot\right)$ is an unknown function of the past returns. 
Here, $\varepsilon_{t}$, for any $t=1,2,\dots,T$, are independent random variables which are supposed to have zero $\tau$--th quantile and constant variance. As noted by \cite{engle_manganelli.2004}, $\ell\left(\cdot\right)$ can be interpreted as the News Impact Curve (NIC) introduced by \cite{engle_ng.1993} for ARCH--type models. The form of the function $\ell\left(\cdot\right)$ is one of the most addressed tasks in the risk modeling literature. Indeed, we can recognize different CAViaR models by considering different form of $\ell\left(\cdot\right)$ such as:
\begin{align}
\ell\left(\bbeta,y_{t-1}\right)&=\beta\vert y_{t-1} \vert,  & \textit{Symmetric Absolute Value}
\label{eq:SAV} \\
\ell\left(\bbeta,y_{t-1}\right) & = \beta_1 \left(y_{t-1}\right)_+ +\beta_2 \left(y_{t-1}\right)_-,  & \textit{Asymmetric Slope}
\label{eq:AS} \\
\ell\left(\bbeta y_{t-1}\right) & =\begin{cases}
\beta_1 \vert y_{t-1} \vert, & z_{t} \leq r \\
\beta_2 \vert y_{t-1} \vert, & z_{t} > r, 
\label{eq:TCAV}
\end{cases}
& \textit{Treshold CAViaR},
\end{align}
where $z_t$ is an observed threshold variable that could be exogenous or $y_t$ itself, i.e., $z_t = y_t$; $r$ is the threshold value, typically set equal to zero, i.e., $r=0$ (see \cite{gerlach_etal.2011}). 
Another possible configuration of the CAViaR model is the Indirect Garch(1,1) where in equation (\ref{eq:caviar_model}) we have
\begin{equation}
q_{t,\tau}(\theta) =\left(\omega + \gamma q^2_{t-1,\tau}(\theta) + \ell\left(\bbeta,y_{t-1}\right)\right)^{1/2}
\label{eq:InG}
\end{equation}
and 
\begin{equation}
\ell\left(\bbeta,y_{t-1}\right)=\beta y^2_{t-1}
\end{equation}

From an inferential point of view, the frequentist approach estimate CAViaR through quantile regression methods (\cite{koenker.2005}, \cite{engle_manganelli.2004}) 
by \cite{koenker_basset.1978} i.e. solving the problem:
\begin{equation}
	\min_{\theta} \frac{1}{T} \sum_{i=1}^{T}\rho_{\tau}(y_{t}-q_{t,\tau}(\theta))
	\label{eq:check_function}
\end{equation}
where $\rho_{\tau}(u)\equiv u\left(\tau-\mathbbm{1}_{(u<0)}\right)$ is the well known quantile \textit{check function}.\newline 
The Bayesian approach, instead, relies on the AL distribution assumption as inferential tool to perform the statistical analysis (see, e.g., \cite{yu_moyeed.2001}, \cite{kottas_gelfand.2001}, \cite{kottas_krnjajic.2009}, \cite{sriram_etal.2013}, \cite{bernardi_etal.2015}).\newline
\indent The same structure considered for the CAViaR models is used to build the CARE model defined as:
\begin{align}
y_t & = \mu_{t,\nu}(\theta) + \varepsilon_t \\
\mu_{t,\nu}(\theta) & =\omega + \gamma \mu_{t-1,\nu}(\theta) + \ell\left(\bbeta,y_{t-1}\right),
\label{eq:care_model}
\end{align}
where $\mu_{t,\nu}(\theta)$ is the $\nu$--th expectile of $y_t$ defined as the value that minimizes the function $\mathbb{E}\left[\left\vert\nu - \bbone_{\left(-\infty,\mu_{t,\nu}(\theta)\right)}\left(y_t\right) \right\vert\left(y_t - \mu_{t,\nu}(\theta) \right)^2\right]$. 
In this model $\varepsilon_{t}$, for any $t=1,2,\dots,T$, are independent random variables having zero $\nu$--th expectile and constant variance. The specifications of $\ell\left(\cdot\right)$  in equations \eqref{eq:SAV}--\eqref{eq:TCAV} remain valid to define different CARE models with the Indirect GARCH(1,1) CARE obtained by substituting in (\ref{eq:InG}) $q_{t,\nu}(\theta)$ with $\mu_{t,\nu}(\theta)$. The estimation procedure for the generic expectile is addressed in the frequentist approach by using the Asymmetric Least Square (ALS) estimator as in \cite{newey_powell.1987}, i.e. by minimizing 
\begin{equation}
	\min_{\theta} \frac{1}{T} \sum_{i=1}^{T}r_{\tau}(y_{t}-\mu_{t,\tau}(\theta))
	\label{eq:care_check_function}
\end{equation}
where $r_{\tau}(u)\equiv u^2\left(\tau-\mathbbm{1}_{(u<0)}\right)$ is the expectile \textit{check function}.\newline
In the Bayesian paradigm the literature relies on the use of the AG distribution as working likelihood in order to make inference (see, e.g., \cite{gerlach_wang.2015}, \cite{gerlach_chen.2014}, \cite{wichitaksorn_etal.2014}, \cite{gerlach_etal.2016}, \cite{gerlach_etal.2016b}).\newline
Finally, since the expectile does not give immediately information on quantile and ES another passage is required to obtain VaR and ES from the estimated expectile. In the first case we simply use the expectile as an estimator of the quantile by iteratively searching for the $\nu$--th expectile for which we observe $\tau \%$ observations in--sample below it. This procedure was suggested by \cite{efron.1991} and allows us to obtain the quantile of interest and consequently the associated VaR level. The ES is instead obtained using the one to one mapping between expectile and ES suggested in \cite{taylor_jw.2008} given by:
\begin{equation}
{\rm ES}_t\left(\tau \right) = \left(1 + \frac{\nu}{\left(1 - 2\nu \right)\tau}\right) \mu_{t,\nu}(\theta) - \frac{\nu}{\left(1 - 2\nu \right)\tau}\mathbb{E}\left(y_t\right).
\label{eq:care2es}
\end{equation}
In the next Section we provide a unified risk measure framework structure which encompasse the CAViaR and CARE ones. 
%
\section{Bayesian Conditional Autoregressive Risk Measures}
\label{sec:B_CARM_models}
%
\noindent In this Section we develop a unified Conditional Autoregressive Risk model which encompasses both the CAViaR and the CARE as particular cases by using the Skew Exponential Power (SEP) proposed by \cite{kobayashi.2015}. In particular we show how one of the parameter of the SEP distribution is responsable for the choise of the autoregressive risk model considered.  
We approach this problem from a Bayesian point of view following \cite{bernardi_bottone_petrella.2018},  
producing a unified framework that we call the Bayesian Conditional Autoregressive Risk Measure (B--CARM) class of models. For this reason let's recall the SEP desity function: 
\begin{equation}
f_{\sS\sE\sP}\left(y_t;g_t,\sigma,\tau,\alpha\right)=c^{-1} \times \left\{
\begin{array}{lc}
 \exp\left\{-\left(1-\tau\right)\left(\frac{g_t-y_t}{\sigma}\right)^\alpha\right\}, & {\rm if } \quad y_t < g_t\\
\exp\left\{-\tau \left(\frac{y_t- g_t}{\sigma}\right)^\alpha\right\}, & {\rm if}\quad y_t\geq g_t,
\end{array}
\right.
\label{eq:SEP}
\end{equation}
where $g_t$ is the location parameter and $\tau\in\left(0,1\right)$ is the skewness parameter. In addiction, $\sigma\in \Re^{+}$ and $\alpha\in\left(0, \infty\right)$ are the scale and shape parameters, respectively, while 
$ c = \sigma\Gamma\left(1+\frac{1}{\alpha}\right) \left(\frac{1}{\tau^{\frac{1}{\alpha}}} + \frac{1}{\left(1 - \tau\right)^{\frac{1}{\alpha}}} \right)  $ and $\Gamma\left(\cdot\right)$
 is the complete gamma function. 
It is easy to check that the SEP distribution encompass the AL and the AG distributions by setting $\alpha=1$ and $\alpha=2$, respectively. 
In this way by considering the generic model specification
\begin{align}
y_t & = g_{t,\nu}(\theta) + \varepsilon_t \label{eq:carm0_model} \\
g_{t,\nu}(\theta) & =\omega + \gamma g_{t-1,\nu}(\theta) + \ell\left(\bbeta,y_{t-1}\right),
\label{eq:carm_model}
\end{align}
with 
\begin{align}
\ell\left(\bbeta,y_{t-1}\right)&=\beta\vert y_{t-1} \vert,  & \textit{Symmetric Absolute Value}
\label{eq:SAV_carm} \\
\ell\left(\bbeta,y_{t-1}\right) & = \beta_1 \left(y_{t-1}\right)_+ +\beta_2 \left(y_{t-1}\right)_-,  & \textit{Asymmetric Slope}
\label{eq:AS_carm} \\
\ell\left(\bbeta y_{t-1}\right) & =\begin{cases}
\beta_1 \vert y_{t-1} \vert, & z_{t} \leq r \\
\beta_2 \vert y_{t-1} \vert, & z_{t} > r, 
\label{eq:TCAV_carm}
\end{cases}
& \textit{Treshold CAViaR},
\end{align}

and 
\begin{equation}
g_{t,\tau}(\theta) =\left(\omega + \gamma g^2_{t,\tau}(\theta) +\ell\left(\bbeta,y_{t-1}\right)\right)^{1/2}
\label{eq:InG_carm}
\end{equation}
where 
\begin{equation}
\ell\left(\bbeta,y_{t-1}\right)=\beta y^2_{t-1}
\label{eq:InG_l}
\end{equation}

and using the SEP distribution as working likelihood we retrieve the CAViaR models by fixing $\alpha=1$ and the CARE ones by fixing $\alpha=2$. Indeed if we use $\alpha=1$ in (\ref{eq:SEP}) than $g_t$ becomes the quantile, while with $\alpha=2$ the $g_t$ becomes the expectile. In this way we are able to build a unified class of autoregressive risk models by using the SEP distribution.
%
As said before, since the expectile does not represent a risk measure itself, when $\alpha=2$ we estimate the VaR by iteratively searching for the expectile for which we observe a given percentage of observations below it, while the ES is obtained using the one to one mapping between expectile and ES given in equation \eqref{eq:care2es}.
\section{Nonlinear CARM}
\label{sec:BNL_CARM_model}
%
\noindent To keep as general as possible the model specified in equations \eqref{eq:carm0_model}--\eqref{eq:InG_l} we present a new Bayesian Non--linear CARM (BNL--CARM) model where in \eqref{eq:carm0_model}--\eqref{eq:carm_model} we set up a nonlinear and semiparametric framework by using a spline approach to model the function $\ell\left(\cdot\right)$. In this way we avoid to impose an apriori functional form for $\ell\left(\cdot\right)$ like the ones considered in the previous sections allowing the data to decide about its shape. In particular we use the B--spline functions of order $d$ with $k$ knots as: 
%
%
\begin{equation}
\ell\left(\bbeta,y_{t-1}\right)=\sum_{\nu=1}^{k+d} \beta_{\nu} B_{\nu} \left( y_{t-1} \right).
\label{eq:bnl_carm}
\end{equation}
\noindent where $B_{\nu} \left( y_{t-1} \right)$ denote the B--spline basis functions and $\beta_{\nu}$ are unknown coefficients to be estimated. 
As it is well known in equation \eqref{eq:bnl_carm} the value of the estimated coefficients and the shape of the fitted function depend upon the number and the position of the knots. In this paper, in absence of specific prior information we assume the equidistance of the knots. Moreover in order to capture the smoothness of the data it is important to carefully choose the number of knots to take in two account of the trade--off between too few and too many knots, which may cause underfitting or overfitting respectively. 
A possible solution of this problem is known as Penalized Spline (P--Spline) and proposed by (\cite{osullivan.1986} and \cite{osullivan.1988}) and generalized by \cite{eilers_marx.1996}. To ensure enough flexibility without incurring in the overfitting problem we use a relatively large number of knots jointly with a penalization term able to smooth sufficiently the fitted curve following \cite{eilers_marx.1996}. In more details we consider a penalty element on the second differences of the B--Spline coefficients which can be embedded in a Bayesian framework by using a second order random walk type prior for all the B--Spline coefficients following \cite{lang_brezger.2004}, \cite{brezger_lang_2006}, \cite{brezger_steiner.2008} and  \cite{bernardi_bottone_petrella.2018} i.e: 
\begin{equation}
\beta_{\nu} = 2\beta_{\nu-1} - \beta_{\nu-2} + u_{\nu},\quad \forall\nu=1,2,\dots,k+d.
\label{eq:randomwalk}
\end{equation}
Here the generic stochastic component $u_{\nu}$ has a Gaussian distribution with zero mean and variance equal to $\phi^2$, i.e. $u_{\nu}\sim\mathcal{N}\left(0,\phi^2\right)$ and $\beta_{\nu-1}$ and $\beta_{\nu-2}$ are initialized with diffuse priors (i.e. $\propto 1$). The smoothness of the fitted curve is controlled by the variance of the error term which correspond to the inverse of the penalization parameter used by \cite{eilers_marx.1996} in the frequentist framework. We choose a conjugate Inverse Gamma prior for $\phi^2$, that is $\phi^2 \sim \mathcal{IG} \left(a^{\left(\phi\right)}, b^{\left(\phi\right)} \right)$ with $a^{\left(\phi\right)}=b^{\left(\phi\right)} = 0.001$. Different choices of hyper parameters are allowed but they all bring to very similar results. Finally, it is possible to write the prior distribution for the B--Spline coefficients as
\begin{equation}
\pi\left(\bbeta \mid\phi\right)\propto \mathcal{N}_{k+d} \left(0,\phi^2 \left(D_2^\prime D_{2}\right)^{-1}\right),
\end{equation}
where $\bbeta=\left(\beta_{1},\beta_{2}\dots,\beta_{k+d}\right)^\prime$, $D_2$ is the difference matrix of dimension $\left(k+d-2\right)\times \left(k+d\right)$ i.e. the differential order of the random walk in equation \eqref{eq:randomwalk} (see also \cite{bernardi_bottone_petrella.2018})
%
\section{Bayesian inference}
\label{sec:Bay_methods}
%
\noindent Bayesian inference requires the specification of the likelihood function as well as the prior distribution for all the parameters of interest. The likelihood function of the model, based on the SEP distribution showed in Section \ref{sec:B_CARM_models} is given by:
\begin{align}
\label{eq:sep_likelihood}
\mathcal{L_{\tau}}(\theta,\sigma & \mid \alpha, \by) =   \left[\sigma\Gamma\left(1+\frac{1}{\alpha}\right) \left(\frac{1}{\tau^{\frac{1}{\alpha}}} + \frac{1}{\left(1 - \tau\right)^{\frac{1}{\alpha}}} \right)\right]^{-T} \times \\
& \prod_{t=1}^T  \left[ \exp\left\{-\left(1-\tau\right)\left(\frac{g_t-y_t}{\sigma}\right)^\alpha\right\} \bbone_{\left(-\infty,g_t\left(\theta\right)\right)}\nonumber+ \exp\left\{-\tau \left(\frac{y_t- g_t}{\sigma}\right)^\alpha\right\}  \bbone_{\left(g_t\left(\theta\right),\infty\right)}\right],
\label{eq:sep_likelihood}
\end{align}
where the vector $\by=(y_1,\ldots,y_T)$ is the sample of observations. The parameter $\alpha$ is fixed equal to $\alpha=1$ or $\alpha=2$, depending on the model we want to estimate, i.e. the quantile model or the expectile one, as showed in previous section.
Concerning the prior specification, we assume the following hierarchical prior structure independent on the value of $\tau$:
\begin{equation} 
\pi\left(\boldsymbol{\Xi}\right)=\pi\left(\boldsymbol{\beta}\mid\phi^2\right)\pi\left(\phi^2\right) \pi\left(\omega\right)\pi\left(\gamma\right)\pi\left(\sigma\right),
\label{eq:prior_distribution}
\end{equation}
with  
\begin{align}
\pi\left(\bbeta\mid\phi^2\right) &\propto \mathcal{N}_{k+d} \left(0,\phi^2 \left(D_2^\prime D_{2}\right)^{-1}\right),
\label{eq:prior_distribution_beta} \\
\pi\left(\phi^2\right) &\propto \mathcal{IG} \left(a^{\left(\phi\right)}, b^{\left(\phi\right)} \right)
\label{eq:prior_distribution_phi}\\
\pi\left(\gamma\right)&\propto \mathcal{N}\left(0,\sigma_\gamma^2 \right)
\label{eq:prior_distribution_gamma}\\
\pi\left(\omega\right)&\propto \mathcal{N}\left(0,\sigma_\omega^2 \right)
\label{eq:prior_distribution_omega}\\
\pi\left(\sigma\right)&\propto\mathcal{IG}\left(a, b\right),
\label{eq:prior_distribution_sigma}
\end{align}
where $\boldsymbol{\Xi}= \left(\theta, \phi^2, \sigma\right)$, $\left(a^{\left(\phi\right)}, b^{\left(\phi\right)}, \sigma_\gamma^2, \sigma_\omega^2, a, b\right)$ are given positive hyperparameters, while $\mathcal{N}\left(\cdot\right)$ and $\mathcal{IG}\left(\cdot\right)$ denote the Normal and the Inverse Gamma distributions respectively.
%
\subsection{The Adaptive Independent Metropolis within Gibbs sampler}
\label{sec:aims_sampler}
%
\noindent The Bayesian inference is carried out using an adaptive MCMC sampling scheme, similar to the one proposed by \cite{bernardi_bottone_petrella.2018}, based on the following posterior distribution
\begin{equation}
\pi\left({\bXi}\mid \by \right) \propto \mathcal{L_{\tau}}\left(\theta,\sigma \mid \alpha, \by \right)\pi\left(\boldsymbol{\beta}\mid\phi^2\right)\pi\left(\phi^2\right) \pi\left(\omega\right)\pi\left(\gamma\right)\pi\left(\sigma\right),
\end{equation}
where $\mathcal{L_{\tau}}\left(\theta,\sigma \mid \alpha, \by \right)$ indicates the likelihood function specified in equation \eqref{eq:sep_likelihood}. 

After choosing a set of initial values for the parameter vector $\bXi^{(0)}$,
the block--move Independent Metropolis within Gibbs (IMG) proceeds by iteratively
simulate candidate values from a given set of proposal distributions which are subsequently accepted o rejected according to the usual Metropolis--Hastings acceptance rule, to preserve the detailed balance condition. Therefore, at the $i$--th iteration, for $i=1,2,\dots$ the simulation algorithm requires a proposal distribution for the parameters $\left(\theta, \sigma\right)$.
Specifically, we consider the following set of proposal distributions
\begin{align}
q\left(\bbeta_{i-1},\bbeta^*_{i}\right) &\sim \mathcal{N}^{k+d}\left(\bmu^{\left(i\right)}_{\beta}, \bSigma^{\left(i\right)}_{\beta} \right)
\label{eq:IMH_propo_beta}\\
q\left(\omega_{i-1},\omega^*_i\right) &\sim \mathcal{N} \left(\mu^{\left(i\right)}_{\omega}, \psi^{\left(i\right)}_{\omega} \right)
\label{eq:IMH_propo_omega}\\
q\left(\gamma_{i-1},\gamma^*_i\right) &\sim \mathcal{N} \left(\mu^{\left(i\right)}_{\gamma}, \psi^{\left(i\right)}_{\gamma} \right)
\label{eq:IMH_propo_gamma}\\
q\left(\sigma_{i-1},\sigma^*_i\right) &\sim \mathcal{N}_{\widetilde{\sigma}}\left(\mu^{\left(i\right)}_{\widetilde{\sigma}}, \psi^{\left(i\right)}_{\widetilde{\sigma}} \right)1/\sigma^*_i
\label{eq:IMH_propo_sigma}
\end{align}
where the scale parameter $\tilde{\sigma}=\log\left(\sigma\right)$  is transformed on the logarithmic scale and subsequently transformed back to preserve positiveness. The Jacobian term in equation \eqref{eq:IMH_propo_sigma} is then required to account for the distribution of the deterministic logarithmic transformation of $\sigma$.

At the $i$--th iteration, the IMG algorithm draws a candidate parameter $\Upsilon^{*}=\left(\xi_1^{*},\xi_2^{*},\xi_3^{*},\xi_4^{*}\right)=\left(\boldsymbol{\beta}^{*},\omega^{*},\gamma^{*},\sigma^{*}\right)$ from \eqref{eq:IMH_propo_beta}--\eqref{eq:IMH_propo_sigma},
which is subsequently accepted or rejected, with probability

\begin{equation}
\lambda\left(\xi_j^{\left(i-1\right)},\xi^*_j\right)= \min\left\{1, 
\frac{\mathcal{L}\left(\xi^*_j,\bXi_{-j}^{\left(i-1\right)}\mid\by,\bx\right)}{\mathcal{L}\left(\bXi^{\left(i-1\right)}\mid\by,\bx\right)}\frac{\pi\left(\xi^*_j\right)}{\pi\left(\xi_j^{\left(i-1\right)}\right)}\frac{q\left(\xi_j^{\left(i-1\right)}\right)}{q\left(\xi^*_j\right)}\right\}, \nonumber
\end{equation}
for $j=1,2,3,4$, where $\lambda\left(\xi_j^{\left(i-1\right)},\xi^*_j\right)$denotes the probability of moving to the new state of the chain, $\pi\left(\cdot\right)$ is the prior given in equations \eqref{eq:prior_distribution_beta}--\eqref{eq:prior_distribution_sigma} and $\bXi_{-j}^{\left(i-1\right)}$ refers to the whole set of parameters at iteration $i-1$ without the $j$--th element of $\Upsilon^{*}$. 
The last step of the algorithm update $\phi^{2}_j$, for $j=1,2,\dots,J$ by simulating directly from the respective full conditional distributions which are proportional to an Inverse Gamma  (\cite{brezger_steiner.2008}) with shape $a^{\left(\phi\right)}_i = a^{\left(\phi\right)}_{i-1} + rank\left(D_{2}^{\prime} D_{2}\right) / 2$ and scale $b^{\left(\phi\right)}_i = b^{\left(\phi\right)}_{i-1} + \beta^{\prime} D_{2}^{\prime} D_{2} \beta$. 
Since most of the statistical properties of the Markov
chain, as well as the performance of the Monte Carlo estimators, depend
crucially  on the definition of the proposal distribution $q\left(\cdot\right)$
(see, e.g., \cite{andrieu_etal.2006} and \cite{andrieu_thoms.2008}), we improve the basic IMG--MCMC algorithm with an additional tuning step that adapts the proposal parameters to the target using the following equations:
\begin{align}
\bmu_\beta^{(i+1)}&=\bmu_\beta^{(i)}+\delta^{(i+1)}\left(\boldsymbol{\beta}-\bmu_\beta^{(i)}\right),\\
\bSigma^{(i+1)}_\beta&=\bSigma^{(i)}_\beta+\delta^{(i+1)} \left( \left(\boldsymbol{\beta}-\bmu_\beta^{(i)}\right)\left(\boldsymbol{\beta}-\bmu_\beta^{(i)}\right)^\trasp - \bSigma^{(i)}_\beta \right),\\
\mu_\omega^{(i+1)}&=\mu_\omega^{(i)}+\varsigma^{(i+1)}\left(\omega-\mu_\omega^{(i)}\right),\\
\psi^{(i+1)}_\omega&=\psi^{(i)}_\omega+\varsigma^{(i+1)} \left( \left(\omega-\mu_\omega^{(i)}\right)^2 - \psi^{(i)}_\omega \right),\\
\mu_\gamma^{(i+1)}&=\mu_\gamma^{(i)}+\varsigma^{(i+1)}\left(\gamma-\mu_\gamma^{(i)}\right),\\
\psi^{(i+1)}_\gamma&=\psi^{(i)}_\gamma+\varsigma^{(i+1)}\left( \left(\gamma-\mu_\gamma^{(i)}\right)^2 - \psi^{(i)}_\gamma \right),\\
\mu_{\widetilde{\sigma}}^{(i+1)}&=\mu_{\widetilde{\sigma}}^{(i)}+\varsigma^{(i+1)}\left(\widetilde{\sigma}-\mu_{\widetilde{\sigma}}^{(i)}\right),\\
\psi^{(i+1)}_{\widetilde{\sigma}}&=\psi^{(i)}_{\widetilde{\sigma}}+\varsigma^{(i+1)} \left( \left(\widetilde{\sigma}-\mu_{\widetilde{\sigma}}^{(i)}\right)^2 - \psi^{(i)}_{\widetilde{\sigma}} \right),
\end{align}

where $\varsigma^{(i+1)}$ denotes a tuning parameter that should be carefully selected in order to ensure the convergence and the ergodicity of the resulting chain (see
\cite{andrieu_etal.2006}). 
\cite{roberts_rosenthal.2007} provide two conditions for the convergence of the chain: the diminishing adaptation condition, which is satisfied if and only if $\varsigma^{(i)}\longrightarrow0$, as $i\rightarrow+\infty$, and the bounded convergence condition,
which  guarantees that all the considered  transition kernels have bounded
convergence time. Moreover, \cite{andrieu_etal.2006} show that both conditions are satisfied if and only if $\varsigma^{(i)}\propto i^{-d}$ where $d\in\left[0.5,1\right]$. Therefore, we choose $\varsigma^{\left(i\right)}=\frac{1}{Ci^{0.5}}$ where  $C=10$. As argued by \cite{roberts_rosenthal.2007}, these two conditions together ensure the asymptotic convergence of the adaptive algorithm and the existence of a weak law of large numbers, respectively.

\section{Empirical applications}
\label{sec:empirical_applications}
To show the usefulness and the simplicity of the unified risk model proposed and to asses its performance on real data we treat five stock market indices: 
Nasdaq (US); Straits Times Index (STI, Singapore); Hang Seng Index (Hong Kong); Corea SE (Corea) and AEX (Holland).
Daily closing stock prices ($P_t$) from January 1, 1988 to November 30, 2018 are obtained from the Thomson Reuters (Datastream) database. The percentage returns are computed as $y_t = (log(P_t)-log(P_{t-1}))\times 100$ and the full sample is divided into an in-sample period (from January 1, 1988 to December 30, 2005) and a forecast (out-of-sample) period of 3250 observations (from January 1, 2006 to November 30, 2018), which covers both the 2008-2009 General Financial Crisis and the following Sovereign Debt Crisis. Summary statistics are displayed in Table \ref{tab:summary_stat}. All the series present the typical stylized facts of financial returns such as positive kurtosis and negative skewness. Moreover, the p-value of the Jarque-Bera test (9-th column) always reject the null hypothesis of normality. 
In the next two subsections we evaluate the forecast ability of the BNL-CAViaR and of the BNL-CARE models i.e. the BNL-CARM when $\alpha=1$ and $\alpha=2$ respectively, for all the series considered and we compare them with the regular CAViaR and CARE specifications.
Tables \ref{tab:backtesting_caviar} and \ref{tab:backtesting_care} show the results of the out-of-sample estimation exercise conduced on all the series specified above.

\subsection{CAViaR forecast evaluation}
\label{sec:forecast_eval_caviar}
As mentioned before the CAViaR models allow to obtain a dynamic VaR estimation. In this section we study the behaviour of the BNL-CAViaR model and of all the CAViaR specification considered in Section \ref{sec:B_CARM_models} comparing the accuracy of the VAR estimation. In particular we label the CAViaR specifications as SAV for the Symmetric Absolute Value, AS for the Asymmetric Slope, T-CAViaR for the Threshold CAViaR and IG for the Indirect Garch.\\ 
The VaR estimation accuracy is evaluated by using the most common backtesting methods based on the comparison between the actual returns with their 1 day ahead VaR forecast.

In Table \ref{tab:backtesting_caviar} we show the results of those tests applied to all the examined series for two $\tau$ levels, i.e.  $\tau=0.01$ and $\tau=0.05$. The first column reports the ratio between the actual and the expected number of violations for a given coverage level $\tau$, i.e. $A/E = \frac{1}{m \tau} \sum_{t=n+1}^{n+m} I\left(y_t < -VaR_t\right) $, where $m=2456$ is the length of the forecast window.
The last three columns report the p-values for common backtesting methods: the unconditional (LR$_{uc}$), the conditional (LR$_{cc}$) coverage tests of \cite{kupiec.1995} and \cite{christoffersen.1998} and the CAViaR Dynamic Quantile (DQ) test of \cite{engle_manganelli.2004} respectively.
The first two are likelihood ratio tests based on the assumption that the hit variable $I_t\left(\tau\right)$ defined as
\begin{equation}
I_t \left(\tau\right) =\left\{
\begin{array}{ll}
1, & {\rm if } \quad y_t < -VaR_{t\mid t-1}\nonumber\\
0, & else\nonumber
\end{array}
\right.
\end{equation}
has a Bernoulli distribution with probability $\tau$.
Under this hypothesis, the LR$_{uc}$ test verifies that the violation probability is equal to the coverage rate, i.e. $P\left(I_t \left(\tau\right)\right) = E\left(I_t \left(\tau\right)\right)=\tau$, while the LR$_{cc}$ test, in addition to LR$_{uc}$ test, also examines the independence hypothesis between violations observed at two different dates. 
The DQ test is instead a regression type test based on the demeaned process associated to $I_t \left(\tau\right)$, namely $H_t\left(\tau\right)= I_t \left(\tau\right) -\tau$. This test uses a regression model to assess the hypothesis of a linear relation between $H_t$, its lagged values and other relevant regressors and is known to be more powerful than the LR$_{uc}$ and the LR$_{cc}$ tests.

\subsection{CARE forecast evaluation}
\label{sec:forecast_eval_care}

While CAViaR models provide an estimate of the $\tau$ level quantile i.e. the risk measure $\tau$-th VaR, CARE models provide an estimate of the $\nu-th$ expectile, which is not a risk measure itself, but can be used to obtain estimates of both the VaR and the ES. Specifically, as anticipated in Section \ref{sec:CARM_models}, from the different CARE specification we derive the $\tau$ level VaR by searching the $\nu-th$ expectile for which the proportion of observations below is $\tau\%$. The value of $\nu$ that satisfies this condition is obtained from the iterative estimation of the CARE models on a grid of step $0.001$ of $\nu$. The $\nu-th$ expectile founded by this recursive searching mechanism, i.e. $\mu_{t,\nu}\left(\theta\right)$, represents an estimate of the $\tau-th$ VaR. Therefore results can be evaluated using the same techniques showed in the previous subsection, i.e. LR$_{cc}$, LR$_{uc}$ and DQ test. The output of these tests, using the specification presented in Section \ref{sec:CARM_models}, are reported in table \ref{tab:backtesting_care}.\\
Moreover, CARE models can be used to deliver estimate of the ES through equation \eqref{eq:care2es} that allows to map the $\nu$-th conditional expectile, $\mu_{t,\nu}\left(\theta\right)$, to the $\tau$-th expected shortfall, $ES_t\left(\tau\right)$.
ES results are more difficult to evaluate and optimal assessment for ES forecasts is still an issue under investigation. 
Here we follow the approach of \cite{mcneil_etal.2005} and \cite{taylor_jw.2008} based on a direct test of the residuals, i.e. the difference between the observations and the ES level for only those observations beyond the quantile VaR prediction. The test assess whether the residuals, standardized by the conditional volatility (the conditional quantile estimate in our case), are i.i.d with zero mean. Specifically, a bootstrap test is implemented (as in \cite{efron_tibshirani.1993} page 224) in order to avoid distributional assumptions. The results of this tests are showed in Table \ref{tab:bootstrap_test_es}.

\subsection{Summary of VaR and ES results}
\label{sec:summary_of_results}

\begin{figure}[ht!]
\begin{center}
\begin{tabular}{cc}
$\tau=0.05$&$\tau=0.01$\\
\multicolumn{2}{c}{\multirow{1}{*}{Nasdaq}} \\
\\
\includegraphics[width=0.2\linewidth]{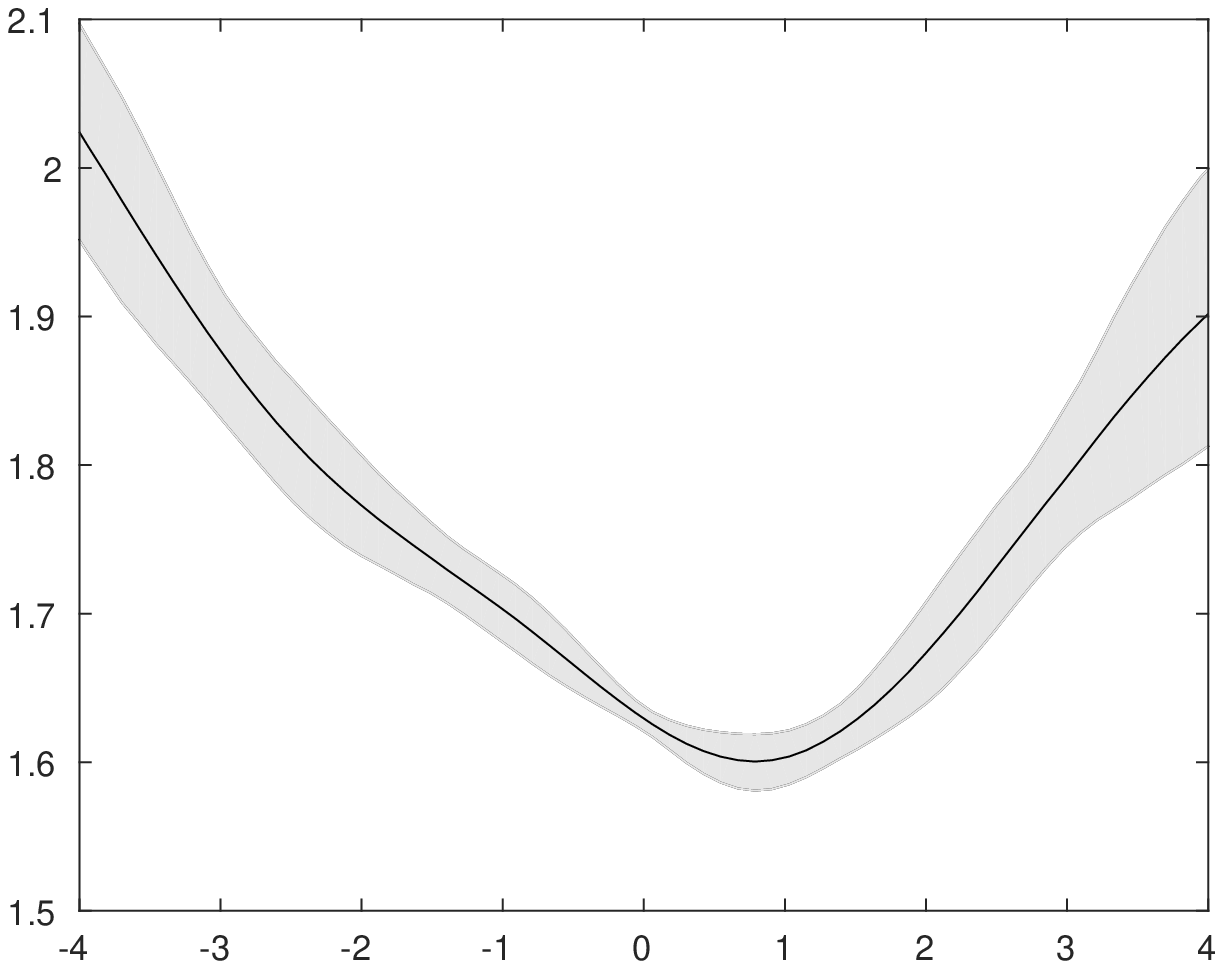}& \qquad
\includegraphics[width=0.2\linewidth]{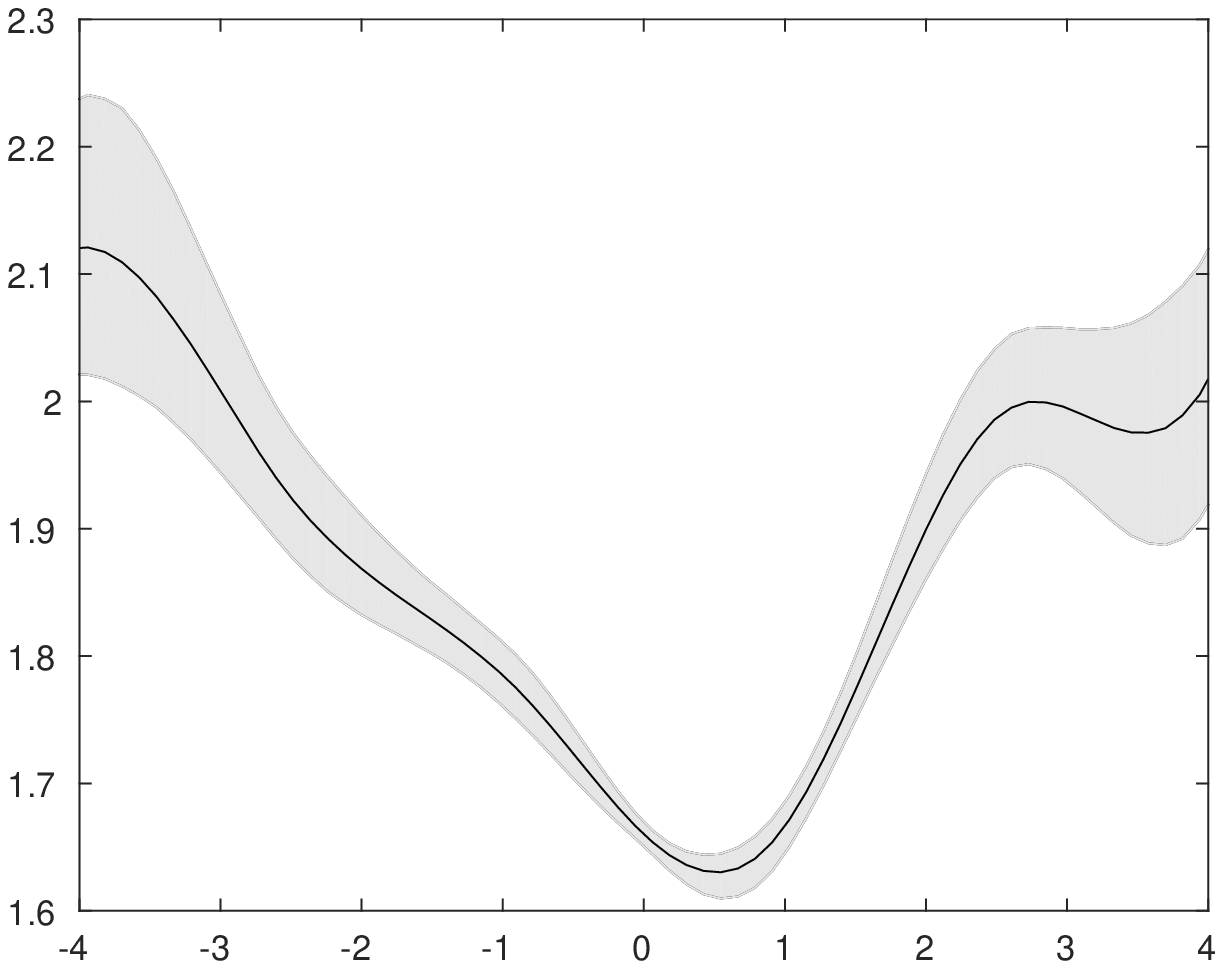}\\
\\
\multicolumn{2}{c}{\multirow{1}{*}{Hang Seng}} \\
\\
\includegraphics[width=0.2\linewidth]{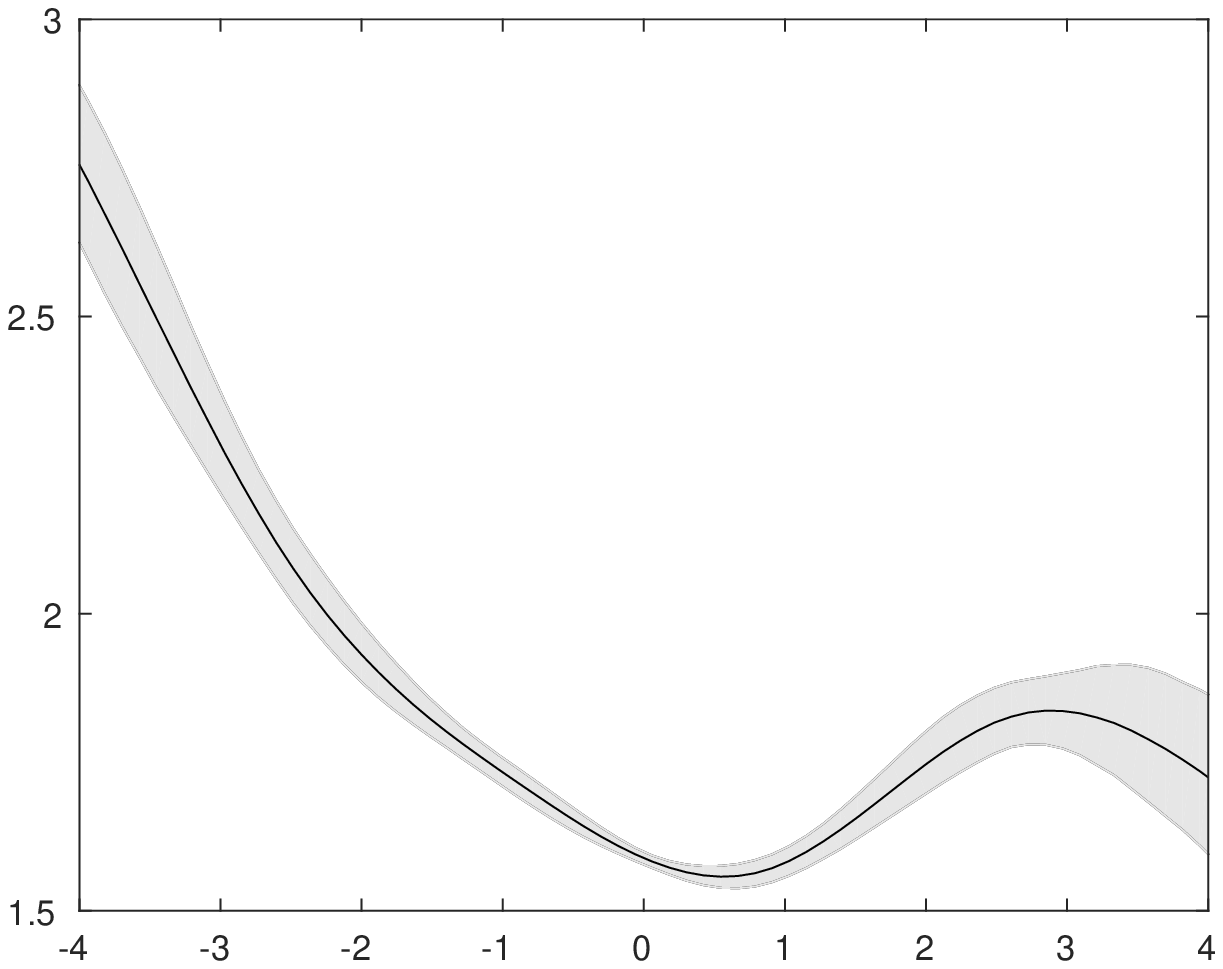}& \qquad
\includegraphics[width=0.2\linewidth]{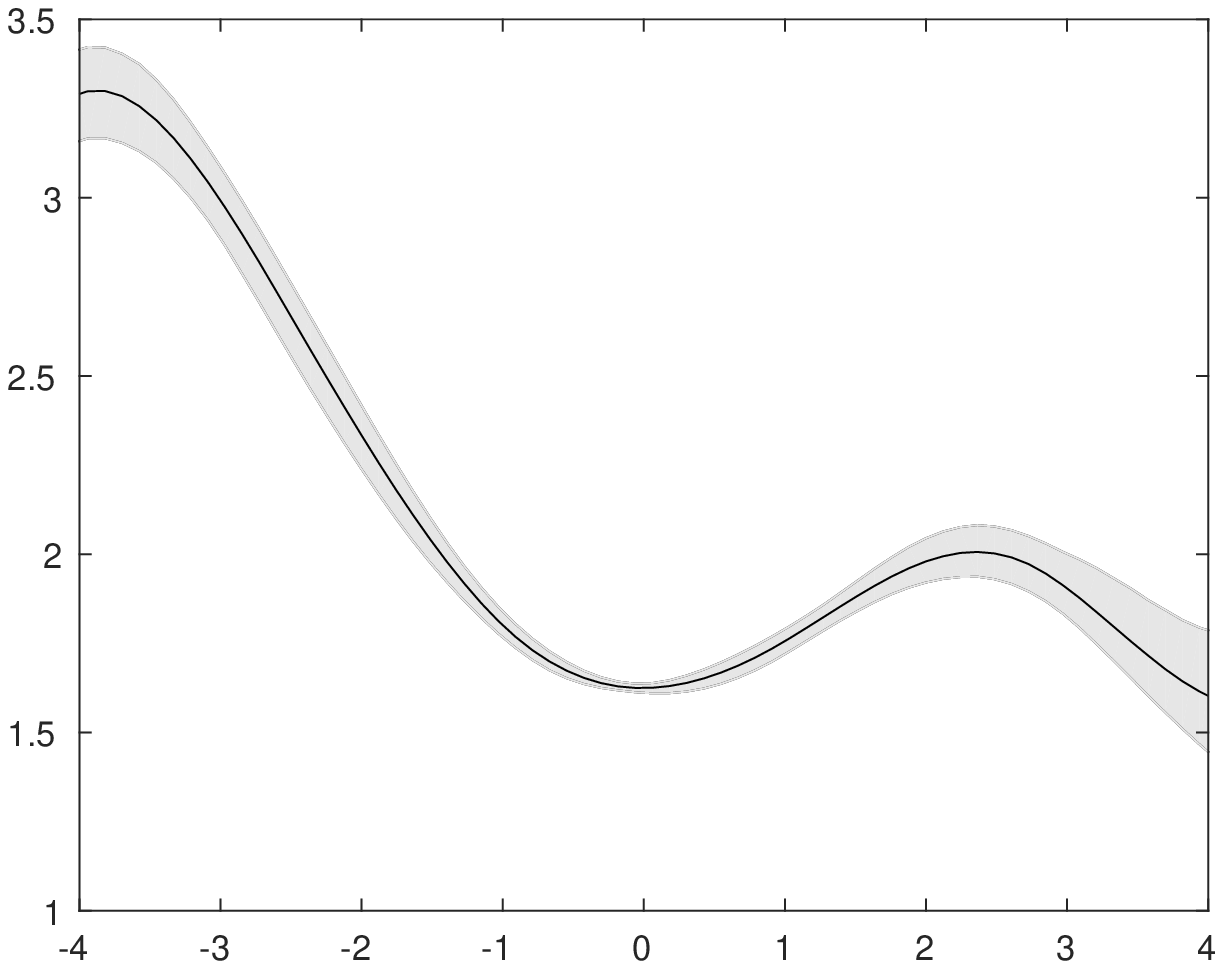}\\
\\
\multicolumn{2}{c}{\multirow{1}{*}{Korea SE}} \\
\\
\includegraphics[width=0.2\linewidth]{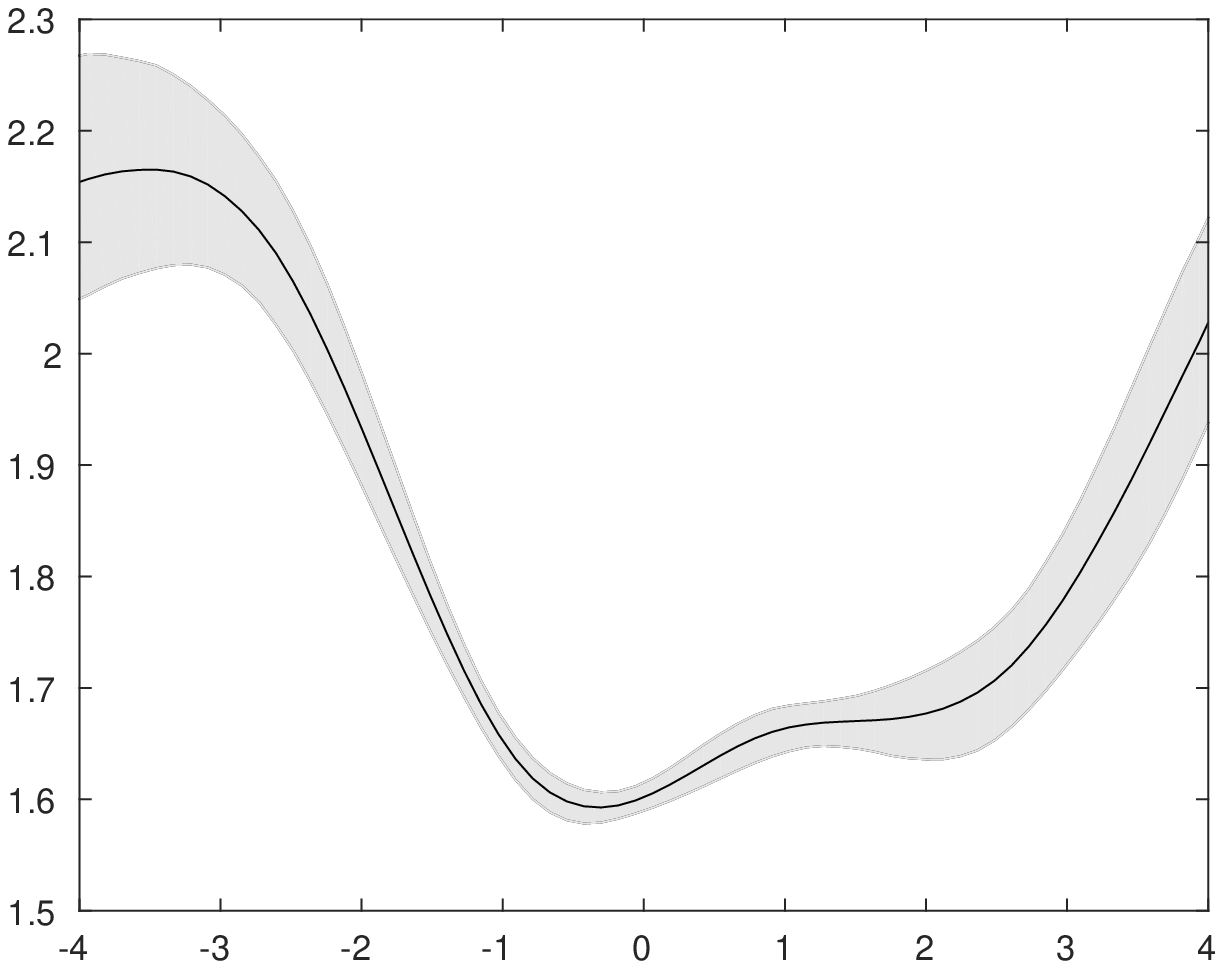}& \qquad
\includegraphics[width=0.2\linewidth]{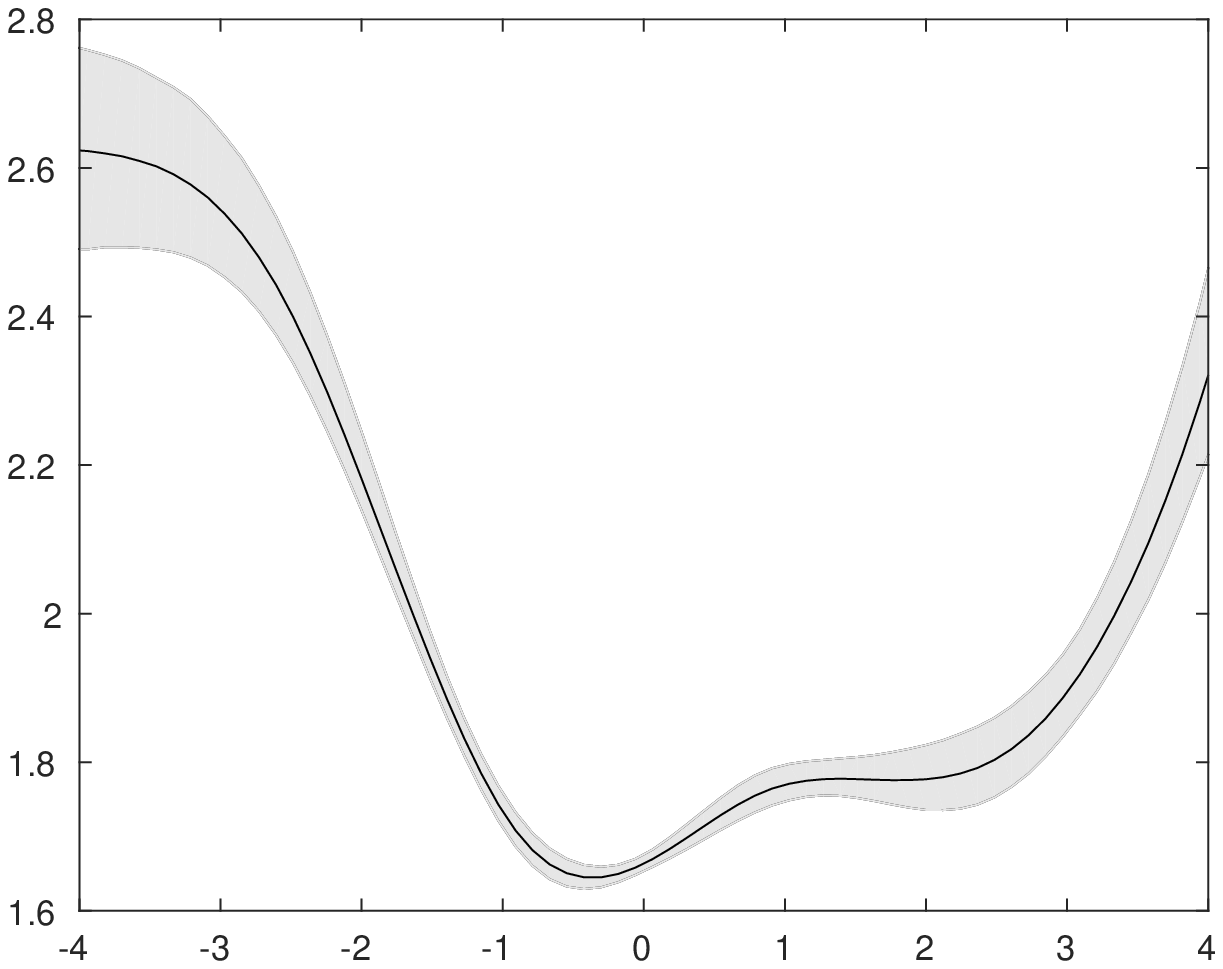}\\
\multicolumn{2}{c}{\multirow{1}{*}{AEX}} \\
\\
\includegraphics[width=0.2\linewidth]{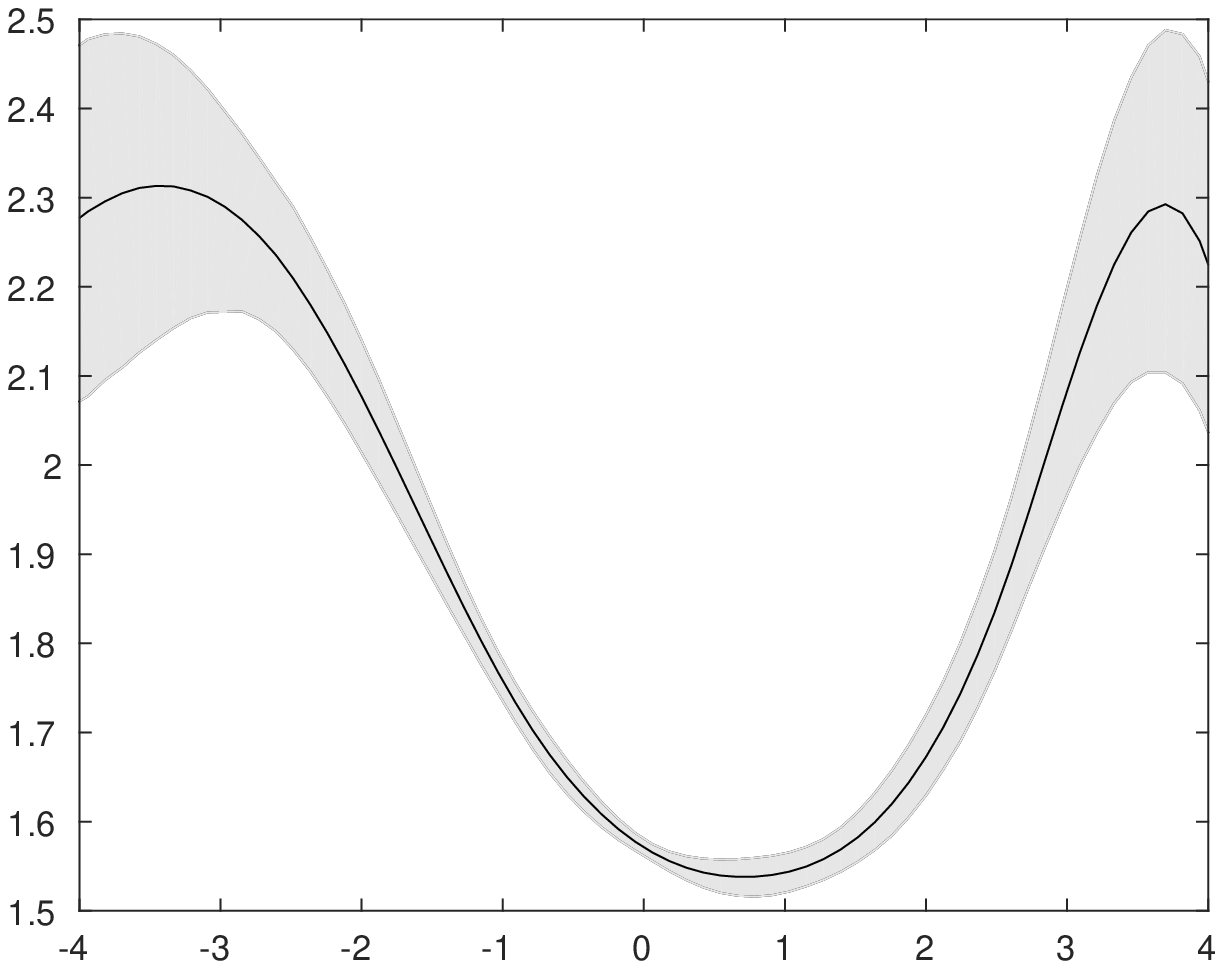}& \qquad
\includegraphics[width=0.2\linewidth]{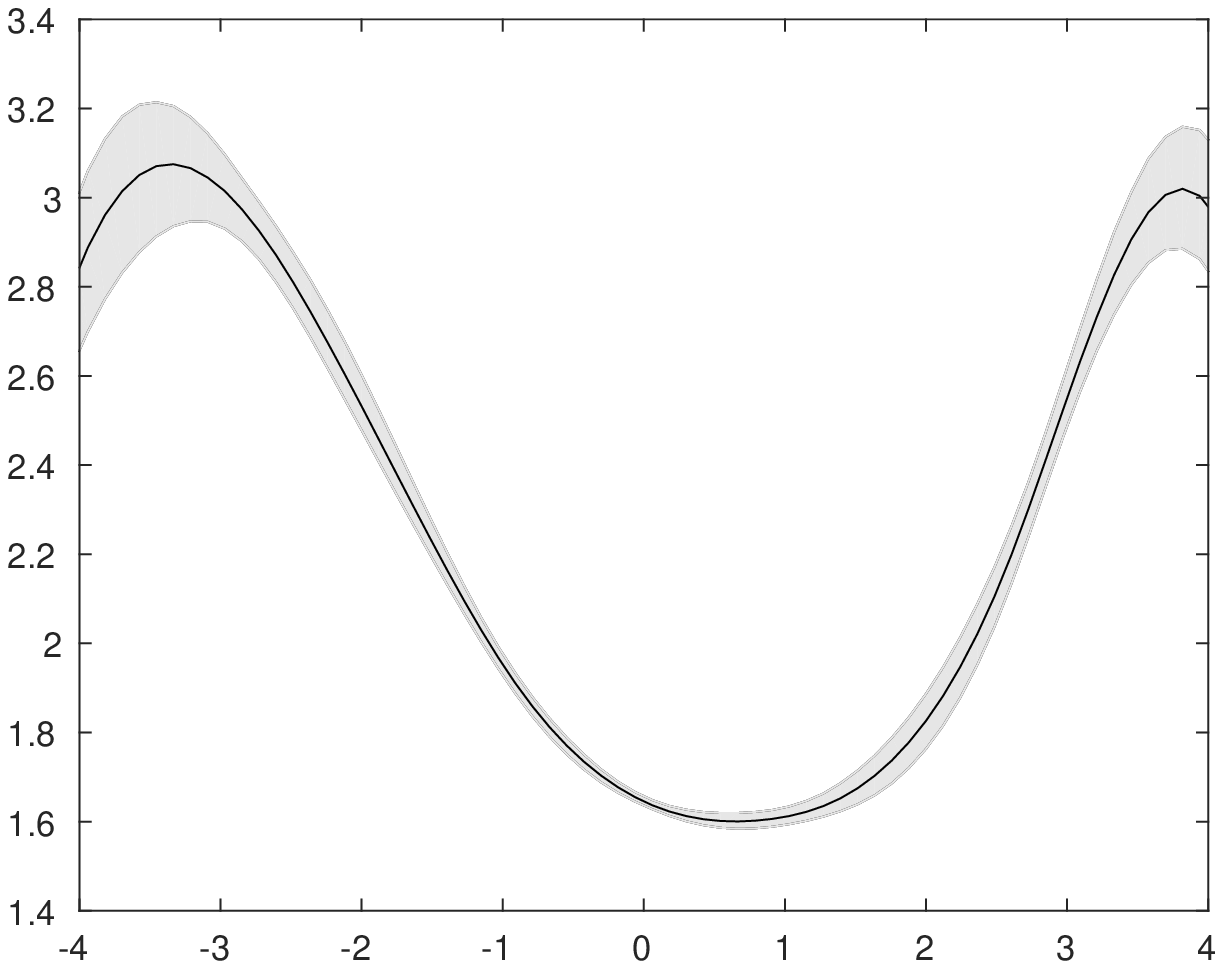}\\
\multicolumn{2}{c}{\multirow{1}{*}{STI}} \\
\\
\includegraphics[width=0.2\linewidth]{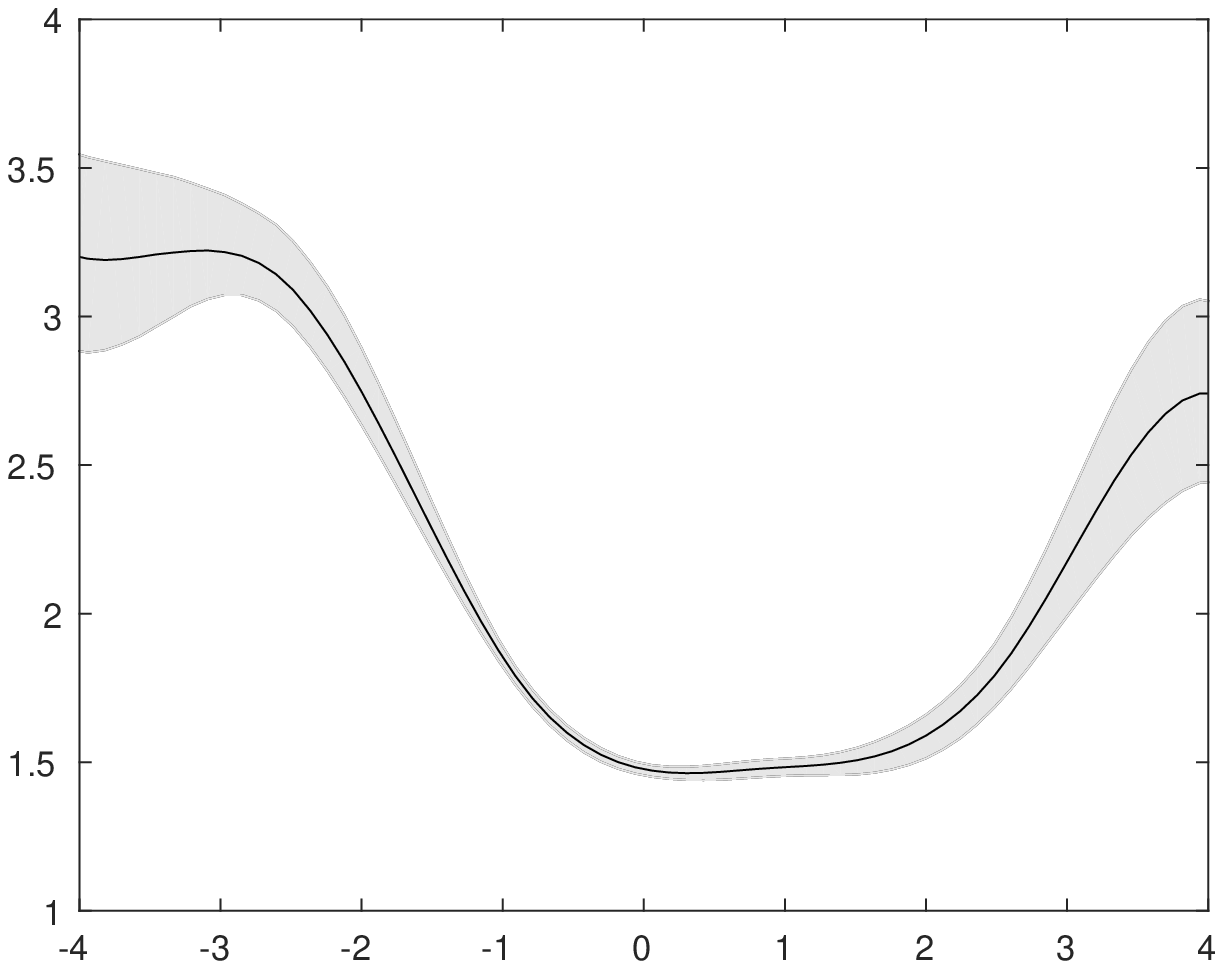}& \qquad
\includegraphics[width=0.2\linewidth]{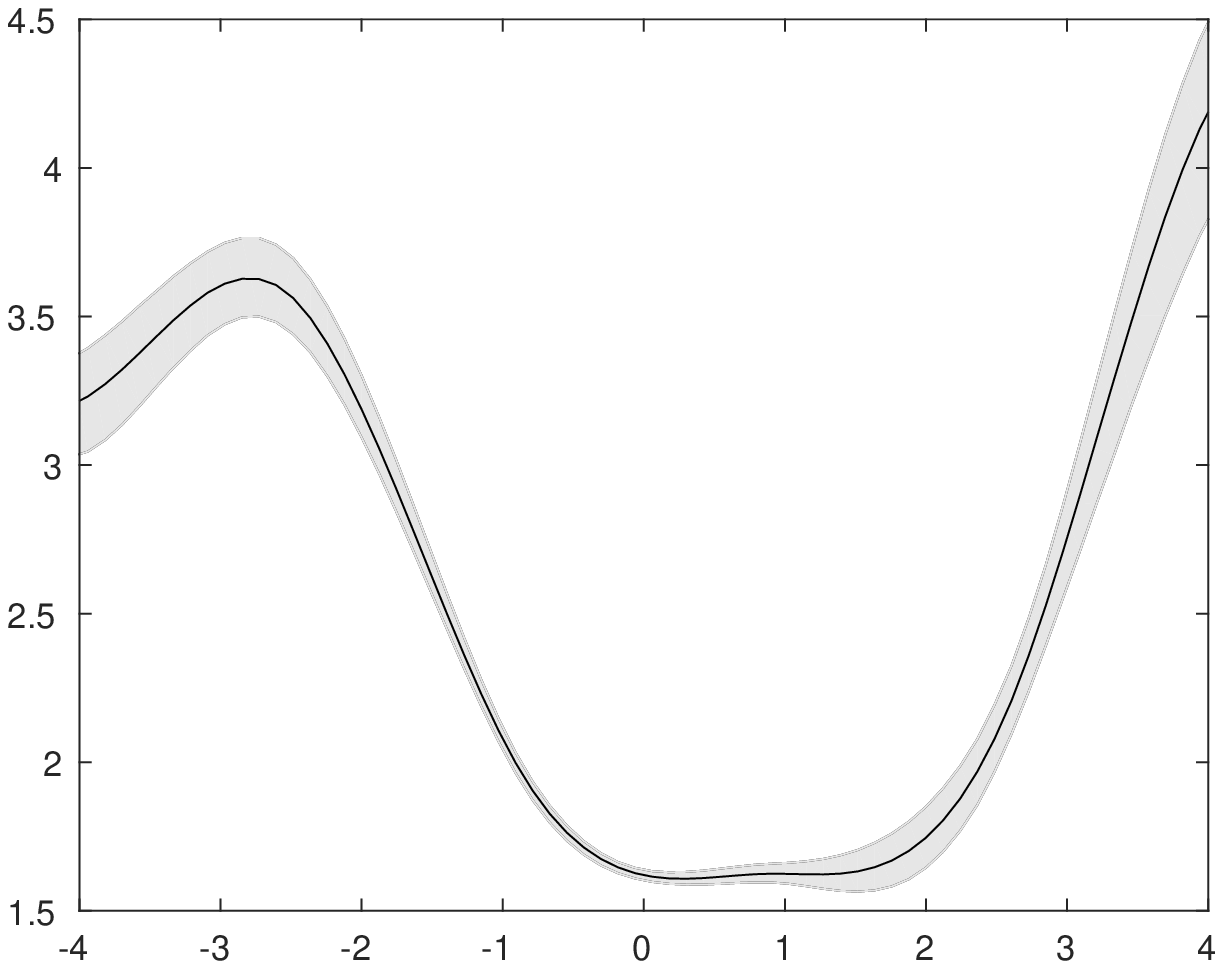}\\
\end{tabular}
\caption{Posterior estimate of the BLN-CAViaR NIC from the empirical application of section \ref{sec:empirical_applications}. Left panels exhibit the NIC (black line) along with the $95\%$ HPD regions (grey areas) at the quantile confidence levels $\tau=\left(0.05\right)$. Right panels exhibit the NIC (black line) along with the $95\%$ HPD regions (grey areas) at the quantile confidence levels $\tau=\left(0.01\right)$.}
\label{fig:NLCAViaR_NIC}
\end{center}
\end{figure}

By analyzing the empirical results given in tables \ref{tab:backtesting_caviar} - \ref{tab:bootstrap_test_es} it is clear that both the BNL-CAViaR and BNL-CARE models show a forecasting performance in line with the other competitor models considered. This can be deducted by looking at the backtesting results reveling that no model outperforms the others in terms of A/E rate, number of violations, or p-values of the LR$_{uc}$, LR$_{uc}$ and DQ tests. By following the traffic light approach suggested by the Basel \cite{basel.1996}, we can say that all the models considered are classified as acceptable (green zone) or at least disputable (yellow zone) and that they never appear to be seriously flawed (red zone).  
It is worth noting that the time lenght of the series considered 
covers both the 2008-2009 General Financial Crisis and the following Sovereign Debt Crisis, events that shocked the market by creating large unexpected loss making the forecast exercise more complex. Despite this situation both BNL-CAViaR and the BNL-CARE perform quite well.\\
In figures \ref{fig:NLCAViaR_NIC} and \ref{fig:NLCARE_NIC} we report the posterior estimate of the NICs from the BNL-CAViaR (BNL-CARM with $\alpha=1$) and BNL-CARE (BNL-CARM with $\alpha=2$) models respectively. For each figure, left panels exhibit the NIC (black line) along with the $95\%$ HPD regions (grey areas) at the quantile confidence levels $\tau=\left(0.05\right)$, while right panels exhibit the NIC (black line) along with the $95\%$ HPD regions (grey areas) at the quantile confidence levels $\tau=\left(0.01\right)$.
\noindent By looking at the tables and at the fugures we can assert that the Spline approach has at least two important advantages despite the bigger number of parameters to be estimated. \\
The first advantages consists to avoid an apriori choice for the shape of the NIC function allowing the data to speak by themself. Indeed all the CAViaR and CARE specifications considered in Section \ref{sec:CARM_models} impose a given 
structure of the NIC providing only a partial estimate of its true shape. The ASV, for istance, impose a "V" shape relation between the risk measure at a given time $t$ and the past observations $y_{t-1}$ allowing just for an estimate of the overall slope of the NIC. The AS model only marginally improve upon the ASV allowing for a different slope for positive and negative returns but still it forces a "V" shape for the NIC function.\\ 
From figures \ref{fig:NLCAViaR_NIC} and \ref{fig:NLCARE_NIC} we can see that the proposed BNL-CARM provides an estimate of the entire shape of NIC giving us unconstrained information on the impact of new information on the risk measure.\\
\noindent Second, the P-Spline approach used to model the NIC also allows us to estimate the threshold level of the returns such that asymmetric response to the risk is observed instead of assuming it equal to zero like in some of the forementioned models (see for example model \eqref{eq:AS} and \eqref{eq:TCAV}). It is well known infact that a common stylized fact of financial time series is that positive and negative returns have a different impact on volatility.
In our model the threshold level is naturally estimated through the NIC function by the data; this point correspond to the (local) minimum of the NIC as reported in figures \ref{fig:NLCAViaR_NIC} and \ref{fig:NLCARE_NIC} and as can be observed is in general slightly greater than zero. This evidence is besides consistent with the recent financial crisis which has increased the degree of risk aversion among traders.

\section{Conclusion}
\label{sec:Conclusion}

\noindent In this paper we present a new extension of the CAViaR and CARE models unified them in a Bayesian quantile regression framework called the B--CARM models by using the SEP likelihood approach. In addiction we propose a semiparametric P--spline framework i.e. the BNL-CARM model to relax some of the shape constrains present in those models in order to have a more flexible way to account for the well known stylized facts about financial time series. 
The Bayesian estimation methodology is carried out using the new adaptive Independent Metropolis--Hastings within
Gibbs (IMG) technique already considered in \cite{bernardi_bottone_petrella.2018}. The results obtained from the model validation backtesting criteria from five stock market indexes show that the model and the estimation methodology effectively capture the nonlinear relation between the unobserved $\tau$--level quantile and its determinants.

\clearpage
\newpage

\appendix
\section{Tables}

\begin{table}[H]
\begin{small}
\begin{center}
\tabcolsep=2.0mm
 \resizebox{1\columnwidth}{!}{%
    \begin{tabular}{lcccccccccc}
    \hline
& Mean & Median & Std. &Min &Max &Q1 &Q3 &Skewness &Normality test &Kurtosis \\
\cmidrule(lr){2-11}
Nasdaq   &0.0488  &0.1135  &1.6542  &-11.1149  &17.2030  &-0.6883  &0.8280  &0.0400  &0.0010  &9.3400  \\
Hang Seng&0.0320  &0.0586  &1.5878  &-24.5202  &17.2471  &-0.6616  &0.7974  &-0.5469  &0.0010  &19.5871  \\
Corea SE &0.0182  &0.0340  &1.6457  &-12.8047  &11.2844  &-0.7179  &0.7682  &-0.1386  &0.0010  &8.5365  \\
AEX      &0.0241  &0.0668  &1.2922  &-9.5903  &10.0283  &-0.5532  &0.6523  &-0.1804  &0.0010  &9.7683  \\
STI      &0.0173  &0.0225  &1.1981  &-10.5446  &12.8738  &-0.5255  &0.5856  &-0.0888  &0.0010  &12.0541  \\
\hline

\end{tabular}}
\caption{\footnotesize{Summary statistics.}}
\label{tab:summary_stat}
\end{center}
\end{small}
\end{table}

\begin{table}[H]
\begin{small}
\begin{center}
\tabcolsep=2.0mm
 \resizebox{1\columnwidth}{!}{%
    \begin{tabular}{lcccccccccccccc}
    \hline
    & \multicolumn{7}{c}{$\tau = 0.01$} & \multicolumn{7}{c}{$\tau = 0.05$}\\
\cmidrule(lr){2-8}\cmidrule(lr){9-15}
\multirow{2}{*}{Name}& \multicolumn{1}{c}{\multirow{2}{*}{A/E}}  & \multicolumn{2}{c}{AD Viol.} & \multicolumn{1}{c}{\multirow{2}{*}{Viol.}} & \multicolumn{1}{c}{\multirow{2}{*}{${\rm LR}_{uc}$}} & \multicolumn{1}{c}{\multirow{2}{*}{${\rm LR}_{cc}$}} & \multicolumn{1}{c}{\multirow{2}{*}{DQ}} &\multicolumn{1}{c}{\multirow{2}{*}{A/E}}  & \multicolumn{2}{c}{AD Viol.} & \multicolumn{1}{c}{\multirow{2}{*}{Viol.}} & \multicolumn{1}{c}{\multirow{2}{*}{${\rm LR}_{uc}$}} & \multicolumn{1}{c}{\multirow{2}{*}{${\rm LR}_{cc}$}} & \multicolumn{1}{c}{\multirow{2}{*}{DQ}} \\
 \cmidrule(lr){3-4}  \cmidrule(lr){10-11}
     & &    Mean & Max &  & &  & &  &  Mean & Max &  & & & \\
\cmidrule(lr){1-1}\cmidrule(lr){2-8}\cmidrule(lr){9-15}

NASDAQ & & & & & & & & & & & & & &\\
BNL-CAViaR&1.1689  &7.4829  &15.5215  &38.0000  &0.3450  &0.1402  &0.0000  &1.0028  &4.9176  &14.6782  &163.0000  &0.9679  &0.9554  &0.0004  \\
SAV       &1.0151  &7.2093  &18.0225  &33.0000  &0.9299  &0.1355  &0.0000  &1.0643  &4.7129  &14.6590  &173.0000  &0.4028  &0.6791  &0.0011  \\
AS        &0.9843  &6.5240  &16.6643  &32.0000  &0.9296  &0.6133  &0.4089  &1.0643  &4.7200  &15.2859  &173.0000  &0.4028  &0.7028  &0.0090  \\
T-CAViaR  &1.4149  &6.0632  &16.4689  &46.0000  &0.0251  &0.0319  &0.0001  &1.1566  &4.6330  &15.3398  &188.0000  &0.0450  &0.1288  &0.0001  \\
IG        &1.0151  &6.9912  &16.2478  &33.0000  &0.9299  &0.1355  &0.0000  &1.0520  &4.6630  &15.1494  &171.0000  &0.4974  &0.7479  &0.0001  \\

HANG SENG & & & & & & & & & & & & & &\\
BNL-CAViaR&0.7221  &8.3603  &23.8684  &23.0000  &0.0976  &0.2145  &0.3949  &1.1177  &5.0657  &23.2981  &178.0000  &0.1332  &0.1884  &0.1971  \\
SAV       &0.6279  &7.0596  &17.5615  &20.0000  &0.0236  &0.0680  &0.1193  &0.9294  &5.3574  &21.8372  &148.0000  &0.3570  &0.4801  &0.1216  \\
AS        &0.5965  &6.5154  &11.3844  &19.0000  &0.0134  &0.0420  &0.3085  &1.1366  &5.1932  &24.4616  &181.0000  &0.0825  &0.0349  &0.0402  \\
T-CAViaR  &0.7849  &8.5494  &18.0606  &25.0000  &0.2056  &0.1916  &0.5401  &1.1680  &5.2449  &25.9787  &186.0000  &0.0336  &0.0045  &0.0052  \\
IG        &0.7849  &7.2136  &17.7435  &25.0000  &0.2056  &0.3682  &0.2197  &1.1115  &5.1525  &21.5238  &177.0000  &0.1547  &0.2192  &0.0692  \\

KOREA SE & & & & & & & & & & & & & &\\
BNL-CAViaR&0.8459  &9.0809  &17.3114  &27.0000  &0.3696  &0.3230  &0.0000  &1.0276  &4.8097  &16.3244  &164.0000  &0.7189  &0.0436  &0.0540  \\
SAV       &0.7519  &8.5358  &21.1337  &24.0000  &0.1411  &0.2822  &0.0922  &1.0965  &4.6451  &17.1827  &175.0000  &0.2163  &0.0013  &0.0000  \\
AS        &0.8459  &7.5243  &17.5978  &27.0000  &0.3696  &0.5310  &0.0136  &1.0965  &4.6490  &17.9097  &175.0000  &0.2163  &0.0609  &0.0831  \\
T-CAViaR  &0.8772  &7.7273  &21.7983  &28.0000  &0.4774  &0.6064  &0.0088  &1.2907  &4.1103  &17.8722  &206.0000  &0.0003  &0.0002  &0.0000  \\
IG        &0.6266  &9.6109  &21.4799  &20.0000  &0.0229  &0.0662  &0.0035  &1.0902  &4.6719  &17.6154  &174.0000  &0.2471  &0.0072  &0.0000  \\

AEX & & & & & & & & & & & & & &\\
BNL-CAViaR&1.1202  &6.6605  &17.4447  &37.0000  &0.4947  &0.5207  &0.7106  &1.0718  &4.5112  &15.4441  &177.0000  &0.3474  &0.4552  &0.6968  \\
SAV       &1.1807  &6.6082  &17.6856  &39.0000  &0.3092  &0.3741  &0.1622  &1.1565  &4.5291  &15.6997  &191.0000  &0.0434  &0.0157  &0.0003  \\
AS        &1.0294  &6.5699  &18.0154  &34.0000  &0.8646  &0.6919  &0.8774  &1.1262  &4.5025  &16.5794  &186.0000  &0.1016  &0.0982  &0.2513  \\
T-CAViaR  &1.1202  &6.3613  &17.7032  &37.0000  &0.4947  &0.5207  &0.8304  &1.1444  &4.5050  &17.1307  &189.0000  &0.0619  &0.0769  &0.1426  \\
IG        &1.3321  &6.4102  &18.6286  &44.0000  &0.0676  &0.1039  &0.1099  &1.1262  &4.4897  &16.4462  &186.0000  &0.1016  &0.0982  &0.0443  \\

STI & & & & & & & & & & & & & &\\
BNL-CAViaR&1.0255  &6.0364  &16.9641  &33.0000  &0.8836  &0.6402  &0.5338  &1.0876  &3.9728  &15.6347  &175.0000  &0.2588  &0.1270  &0.0291  \\
SAV       &1.0566  &5.7375  &12.4956  &34.0000  &0.7480  &0.6605  &0.7565  &1.0193  &3.8977  &15.5282  &164.0000  &0.7995  &0.9429  &0.7876  \\
AS        &1.0255  &5.7592  &13.6595  &33.0000  &0.8836  &0.6402  &0.7410  &1.0690  &3.9577  &16.9818  &172.0000  &0.3722  &0.4894  &0.1332  \\
T-CAViaR  &1.0876  &5.7188  &12.9416  &35.0000  &0.6210  &0.6189  &0.6659  &1.0752  &3.9313  &17.0519  &173.0000  &0.3313  &0.4417  &0.1417  \\
IG        &1.0876  &5.8194  &13.6148  &35.0000  &0.6210  &0.6022  &0.1269  &1.0379  &3.9325  &15.0095  &167.0000  &0.6209  &0.6401  &0.8215  \\

\hline
\end{tabular}}
  \caption{\footnotesize{Summary statistics for CAViaR models. The columns, denoted by LR$_{uc}$, LR$_{cc}$ and DQ, report the p--values of unconditional and conditional coverage tests of Kupiec \cite{kupiec.1995} and Christoffersen \cite{christoffersen.1998} and those of the Dynamic Quantile (DQ) test of Engle and Manganelli \cite{engle_manganelli.2004}.}}
  \label{tab:backtesting_caviar}
    \end{center}
    \end{small}
\end{table}

\begin{table}[H]
\begin{small}
\begin{center}
\tabcolsep=2.0mm
 \resizebox{1\columnwidth}{!}{%
    \begin{tabular}{lcccccccccccccc}
    \hline
    & \multicolumn{7}{c}{$\tau = 0.01$} & \multicolumn{7}{c}{$\tau = 0.05$}\\
\cmidrule(lr){2-8}\cmidrule(lr){9-15}
\multirow{2}{*}{Name}& \multicolumn{1}{c}{\multirow{2}{*}{A/E}}  & \multicolumn{2}{c}{AD Viol.} & \multicolumn{1}{c}{\multirow{2}{*}{Viol.}} & \multicolumn{1}{c}{\multirow{2}{*}{${\rm LR}_{uc}$}} & \multicolumn{1}{c}{\multirow{2}{*}{${\rm LR}_{cc}$}} & \multicolumn{1}{c}{\multirow{2}{*}{DQ}} &\multicolumn{1}{c}{\multirow{2}{*}{A/E}}  & \multicolumn{2}{c}{AD Viol.} & \multicolumn{1}{c}{\multirow{2}{*}{Viol.}} & \multicolumn{1}{c}{\multirow{2}{*}{${\rm LR}_{uc}$}} & \multicolumn{1}{c}{\multirow{2}{*}{${\rm LR}_{cc}$}} & \multicolumn{1}{c}{\multirow{2}{*}{DQ}} \\
 \cmidrule(lr){3-4}  \cmidrule(lr){10-11}
     & &    Mean & Max &  & &  & &  &  Mean & Max &  & & & \\
\cmidrule(lr){1-1}\cmidrule(lr){2-8}\cmidrule(lr){9-15}

NASDAQ & & & & & & & & & & & & & &\\
BNL-CARE&0.7075  &8.5479  &16.6762  &23.0000  &0.0772  &0.0764  &0.0275  &0.8920  &5.3056  &15.0000  &145.0000  &0.1517  &0.3496  &0.0002  \\
SAV     &0.8920  &8.0395  &17.9063  &29.0000  &0.5297  &0.0703  &0.0005  &1.0889  &4.8234  &14.9587  &177.0000  &0.2497  &0.5032  &0.0012  \\
AS      &0.7075  &8.0443  &16.5115  &23.0000  &0.0772  &0.1780  &0.1135  &1.0151  &4.8593  &15.5370  &165.0000  &0.8409  &0.6495  &0.0191  \\
T-CARE  &0.6152  &7.9794  &16.3227  &20.0000  &0.0177  &0.0530  &0.0437  &1.0458  &4.8962  &15.8722  &170.0000  &0.5490  &0.7937  &0.0006  \\
IG      &0.7075  &8.0717  &16.3479  &23.0000  &0.0772  &0.0764  &0.0580  &1.0581  &4.8623  &15.4656  &172.0000  &0.4486  &0.7153  &0.0000  \\

HANG SENG & & & & & & & & & & & & & &\\
BNL-CARE&0.6279  &10.0314  &23.1171  &20.0000  &0.0236  &0.0680  &0.0116  &1.0801  &5.3312  &22.1913  &172.0000  &0.3039  &0.0725  &0.0730  \\
SAV     &0.6593  &6.8444  &17.8665  &21.0000  &0.0396  &0.1046  &0.1401  &0.9922  &5.1728  &23.5544  &158.0000  &0.9222  &0.2856  &0.0689  \\
AS      &0.5651  &6.0058  &10.1083  &18.0000  &0.0073  &0.0246  &0.1670  &1.1303  &5.1605  &25.2195  &180.0000  &0.0973  &0.0832  &0.1786  \\
T-CARE  &0.5024  &8.3517  &12.6012  &16.0000  &0.0018  &0.0071  &0.2169  &1.0612  &5.3918  &15.7235  &169.0000  &0.4299  &0.1099  &0.0259  \\
IG      &0.7849  &7.2146  &17.4942  &25.0000  &0.2056  &0.3682  &0.2180  &1.0487  &5.2958  &22.4331  &167.0000  &0.5291  &0.6650  &0.5228  \\

KOREA SE & & & & & & & & & & & & & &\\
BNL-CARE&0.6892  &8.6710  &19.1366  &22.0000  &0.0618  &0.1500  &0.1472  &0.9837  &4.9451  &16.0867  &157.0000  &0.8355  &0.0464  &0.0507  \\
SAV     &0.7832  &8.5001  &20.7419  &25.0000  &0.2013  &0.3628  &0.0525  &1.0276  &4.8199  &17.1703  &164.0000  &0.7189  &0.0882  &0.0178  \\
AS      &0.7519  &7.8578  &17.6468  &24.0000  &0.1411  &0.2822  &0.0075  &1.0526  &4.7782  &18.0365  &168.0000  &0.4960  &0.0552  &0.0405  \\
T-CARE  &0.8145  &7.7491  &17.6432  &26.0000  &0.2773  &0.4477  &0.0153  &1.2218  &4.3582  &17.7446  &195.0000  &0.0053  &0.0068  &0.0193  \\
IG      &0.7832  &8.7727  &21.1006  &25.0000  &0.2013  &0.3628  &0.0172  &1.0526  &4.8380  &17.3711  &168.0000  &0.4960  &0.0268  &0.0002  \\

AEX & & & & & & & & & & & & & &\\
BNL-CARE&1.0294  &7.5506  &17.3401  &34.0000  &0.8646  &0.6919  &0.1884  &1.1262  &4.6867  &17.7054  &186.0000  &0.1016  &0.1425  &0.0469  \\
SAV     &1.2716  &7.8593  &16.8078  &42.0000  &0.1318  &0.1870  &0.0000  &1.2231  &4.5865  &14.9509  &202.0000  &0.0044  &0.0047  &0.0001  \\
AS      &1.1202  &7.6155  &16.0772  &37.0000  &0.4947  &0.5207  &0.0095  &1.1505  &4.6591  &15.7784  &190.0000  &0.0519  &0.1227  &0.0066  \\
T-CARE  &1.2110  &7.8290  &17.3910  &40.0000  &0.2374  &0.3046  &0.0000  &1.2171  &4.5207  &15.4260  &201.0000  &0.0055  &0.0185  &0.0016  \\
IG      &1.2716  &7.5170  &16.7142  &42.0000  &0.1318  &0.2719  &0.0003  &1.1565  &4.5903  &15.7695  &191.0000  &0.0434  &0.0268  &0.0085  \\

STI & & & & & & & & & & & & & &\\
BNL-CARE&0.9944  &7.1228  &16.9036  &32.0000  &0.9759  &0.6198  &0.0001  &1.2927  &3.7855  &14.9127  &208.0000  &0.0003  &0.0011  &0.0001  \\
SAV     &0.8701  &7.4723  &16.3825  &28.0000  &0.4499  &0.3822  &0.0000  &0.9323  &4.2518  &14.3585  &150.0000  &0.3749  &0.6746  &0.7608  \\
AS      &0.9012  &7.2291  &15.0321  &29.0000  &0.5678  &0.4559  &0.0009  &1.0068  &4.3077  &15.3054  &162.0000  &0.9260  &0.9940  &0.0005  \\
T-CARE  &0.8080  &7.4438  &17.3445  &26.0000  &0.2581  &0.2383  &0.0028  &0.9944  &4.3194  &15.4082  &160.0000  &0.9451  &0.9975  &0.0012  \\
IG      &0.9012  &7.0910  &16.3101  &29.0000  &0.5678  &0.6525  &0.0308  &0.9447  &4.2920  &14.0801  &152.0000  &0.4701  &0.4451  &0.2950  \\
\hline
\end{tabular}}
  \caption{\footnotesize{Summary statistics for CARE models. The columns, denoted by LR$_{uc}$, LR$_{cc}$ and DQ, report the p--values of unconditional and conditional coverage tests of Kupiec \cite{kupiec.1995} and Christoffersen \cite{christoffersen.1998} and those of the Dynamic Quantile (DQ) test of Engle and Manganelli \cite{engle_manganelli.2004}.}}
  \label{tab:backtesting_care}
    \end{center}
    \end{small}
\end{table}

\begin{table}[H]
\begin{small}
\begin{center}
\tabcolsep=2.0mm
 \resizebox{1\columnwidth}{!}{%
    \begin{tabular}{lcccccccccc}
    \hline
& \multicolumn{5}{c}{$\tau = 0.01$} & \multicolumn{5}{c}{$\tau = 0.05$}\\
\cmidrule(lr){2-6}\cmidrule(lr){7-11}
\multirow{1}{*}{Name} &
\multicolumn{1}{c}{\multirow{1}{*}{BNL-CARE}} &  \multicolumn{1}{c}{\multirow{1}{*}{SAV}} 		 & 
\multicolumn{1}{c}{\multirow{1}{*}{AS}}  			&
\multicolumn{1}{c}{\multirow{1}{*}{T-CARE}}		&
\multicolumn{1}{c}{\multirow{1}{*}{IG}}  			& 
\multicolumn{1}{c}{\multirow{1}{*}{BNL-CARE}} &  \multicolumn{1}{c}{\multirow{1}{*}{SAV}} 		 & 
\multicolumn{1}{c}{\multirow{1}{*}{AS}}  			&
\multicolumn{1}{c}{\multirow{1}{*}{T-CARE}}		&
\multicolumn{1}{c}{\multirow{1}{*}{IG}}  			\\ 
\cmidrule(lr){1-1}\cmidrule(lr){2-11}

Nasdaq   &0.5260  &0.9400  &0.8590  &0.2220  &0.7300  &0.0760  &0.1750  &0.5370  &0.6780  &0.3840  \\
Hang Seng&0.0000  &0.0000  &0.0000  &0.8340  &0.0000  &0.0000  &0.0000  &0.0000  &0.0200  &0.0010  \\
Corea SE &0.1310  &0.0080  &0.0080  &0.0440  &0.0150  &0.1090  &0.0000  &0.0000  &0.0080  &0.0050  \\
AEX      &0.0080  &0.0000  &0.0020  &0.0010  &0.0000  &0.0060  &0.0490  &0.0050  &0.0090  &0.4720  \\
STI      &0.0000  &0.0000  &0.0000  &0.0000  &0.0000  &0.0130  &0.0010  &0.0000  &0.0000  &0.0050  \\

\hline
\end{tabular}}
\caption{\footnotesize{Bootstrap test p-values for zero mean of the standardardized residuals. Test based on 1000 post sample estimates of the conditional 1\% and 5\% ES.}}
\label{tab:bootstrap_test_es}
\end{center}
\end{small}
\end{table}

\newpage
\section{Figures}
%
\vspace{-0.5cm}
%

\begin{figure}[ht!]
\begin{center}
\begin{tabular}{cc}
$\tau=0.05$&$\tau=0.01$\\
\multicolumn{2}{c}{\multirow{1}{*}{Nasdaq}} \\
\\
\includegraphics[width=0.2\linewidth]{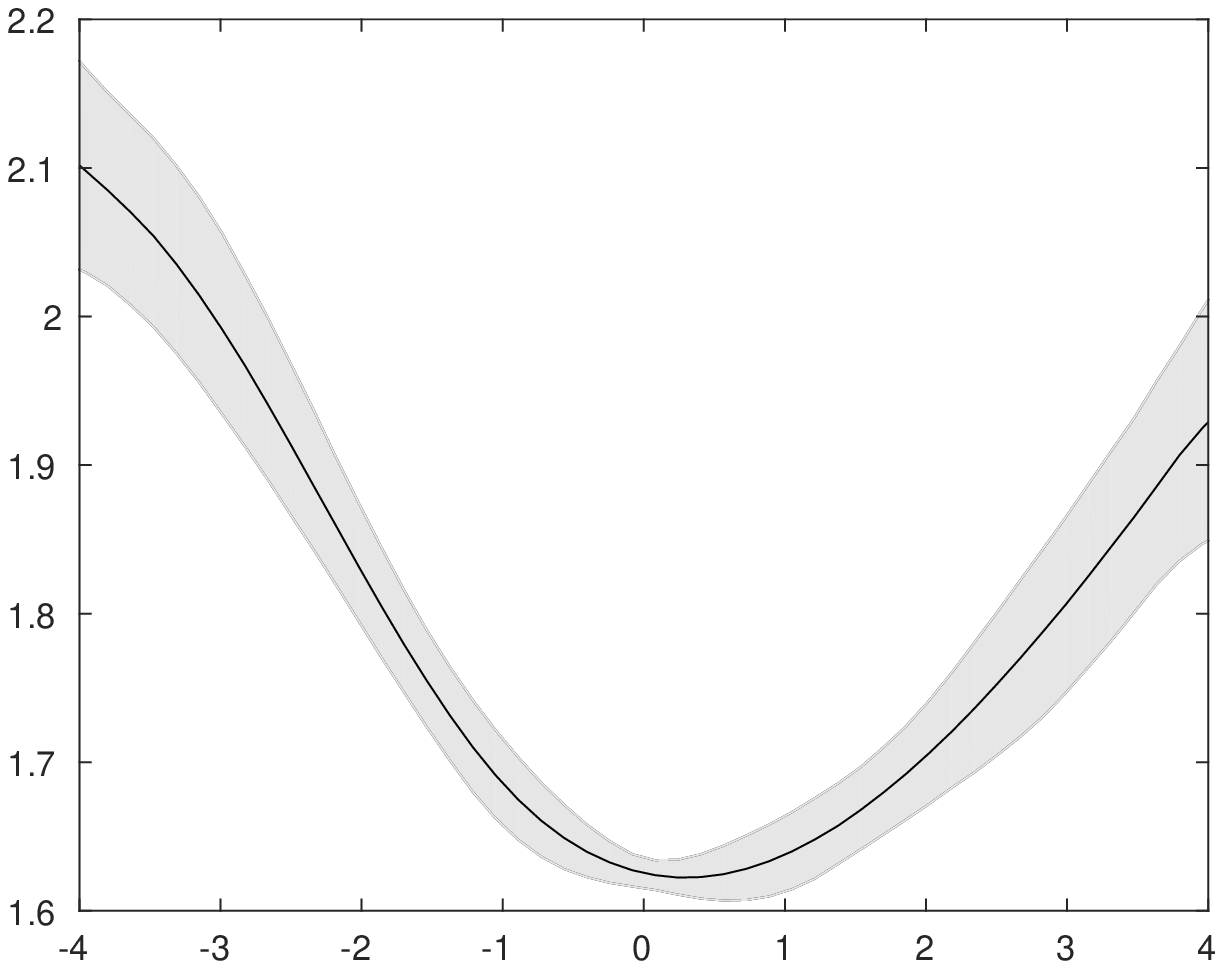}& \qquad
\includegraphics[width=0.2\linewidth]{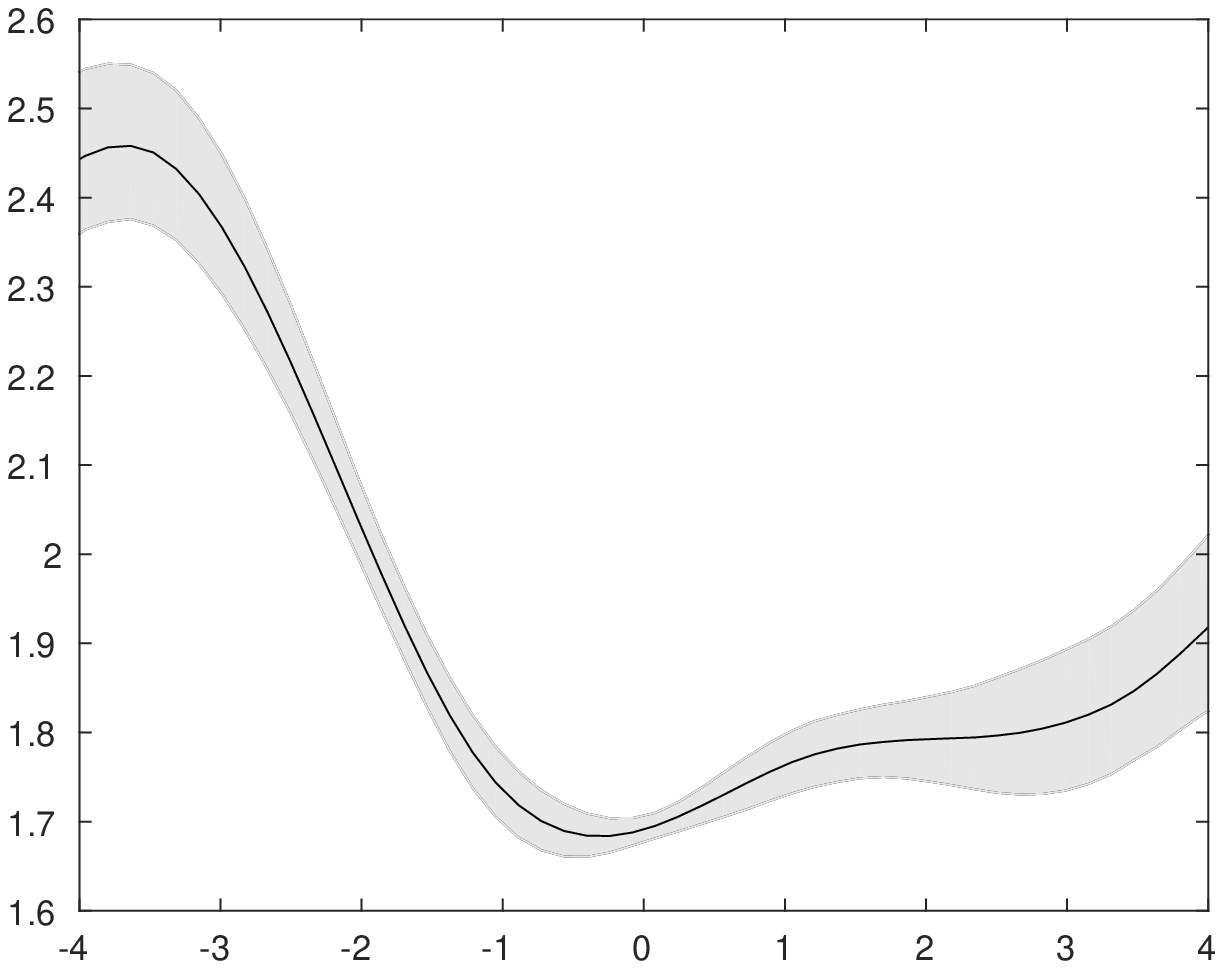}\\
\\
\multicolumn{2}{c}{\multirow{1}{*}{Hang Seng}} \\
\\
\includegraphics[width=0.2\linewidth]{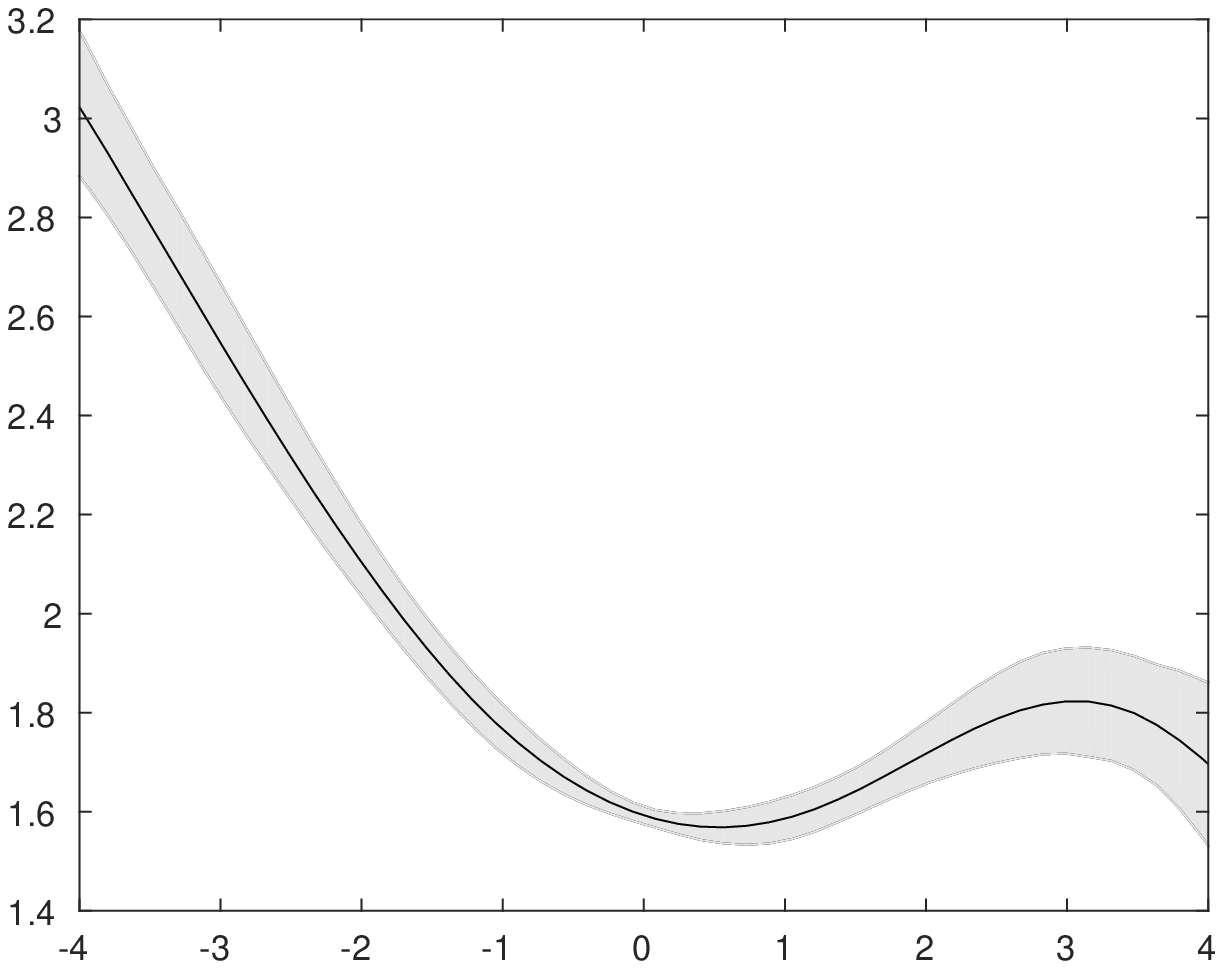}& \qquad
\includegraphics[width=0.2\linewidth]{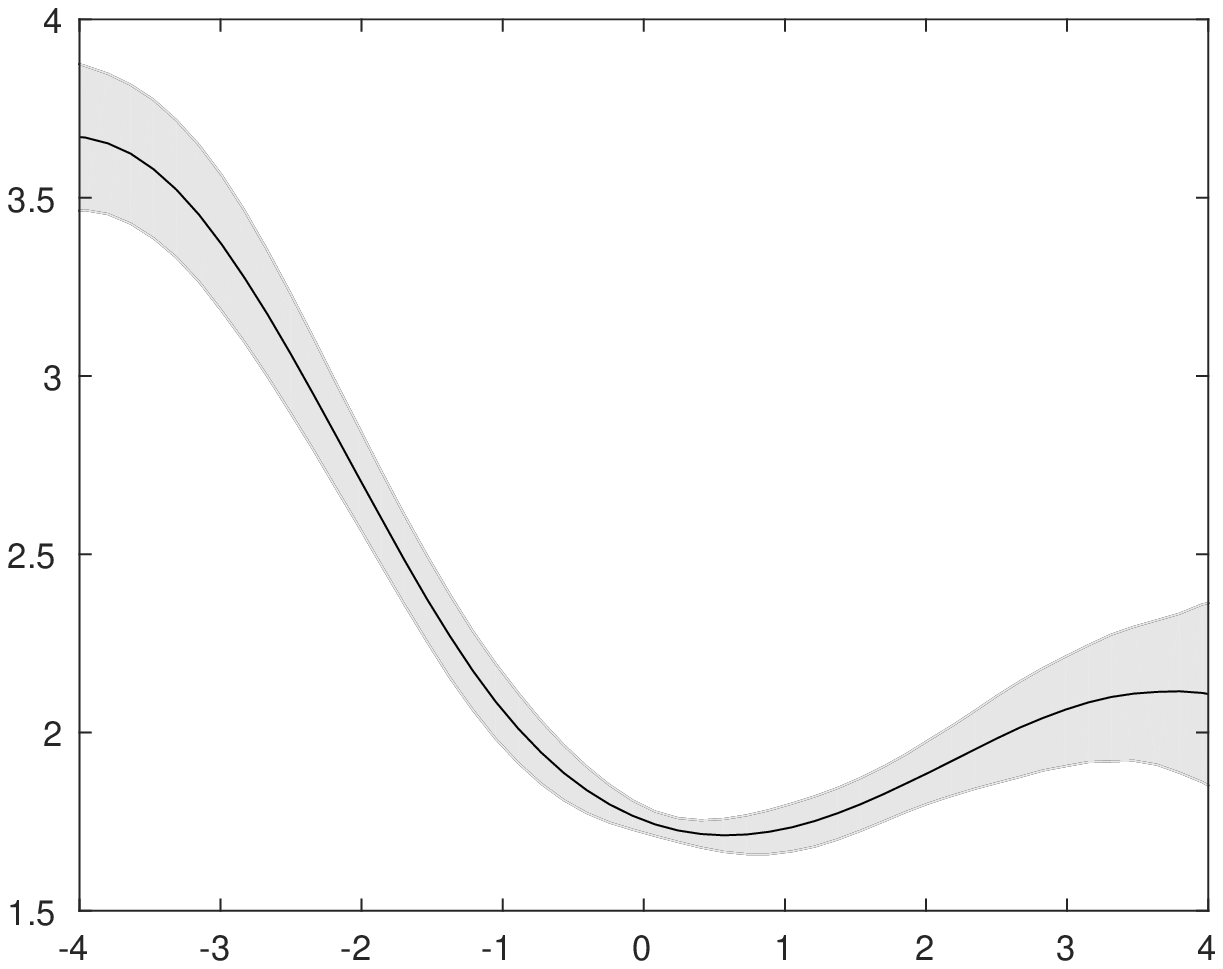}\\
\\
\multicolumn{2}{c}{\multirow{1}{*}{Korea SE}} \\
\\
\includegraphics[width=0.2\linewidth]{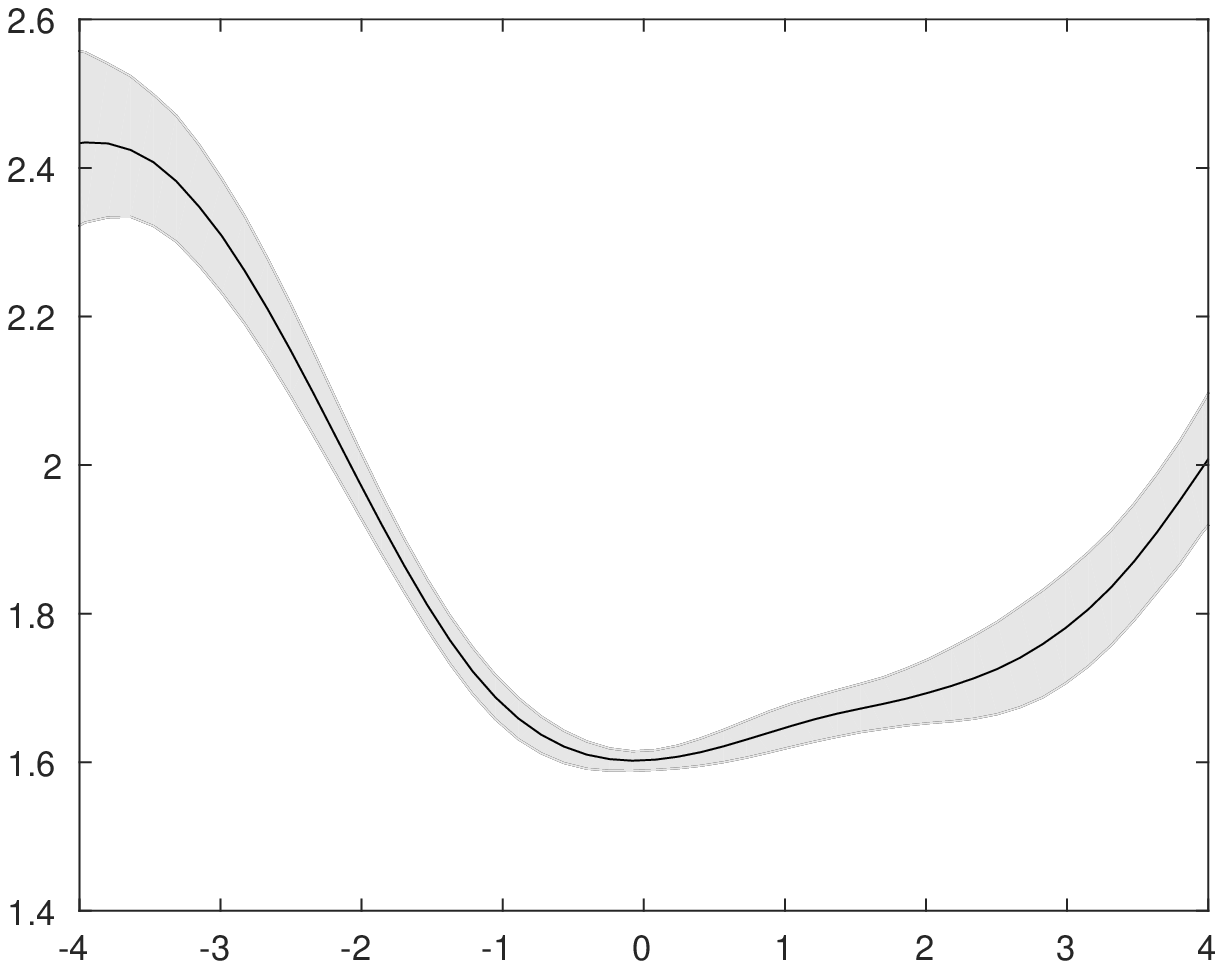}& \qquad
\includegraphics[width=0.2\linewidth]{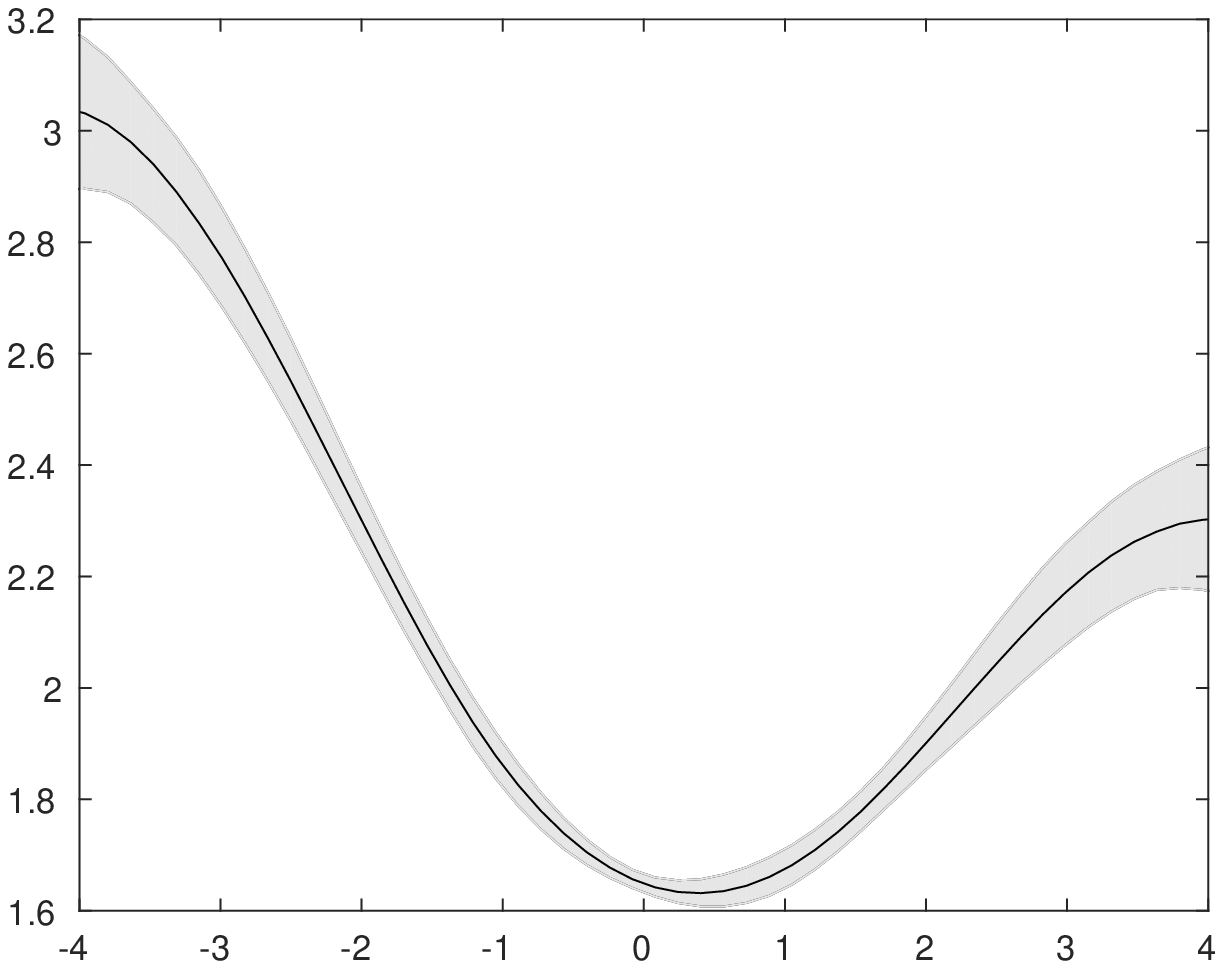}\\
\multicolumn{2}{c}{\multirow{1}{*}{AEX}} \\
\\
\includegraphics[width=0.2\linewidth]{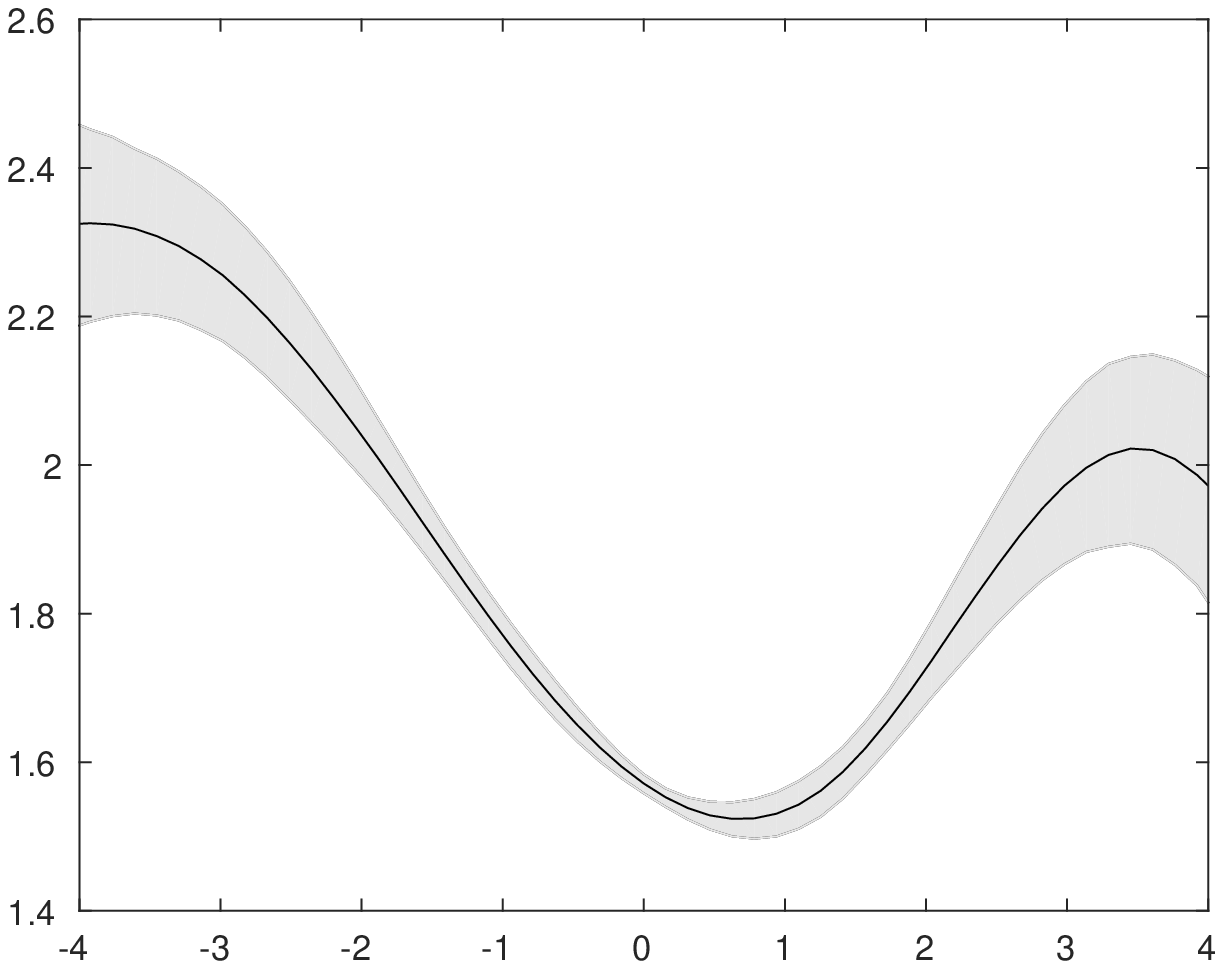}& \qquad
\includegraphics[width=0.2\linewidth]{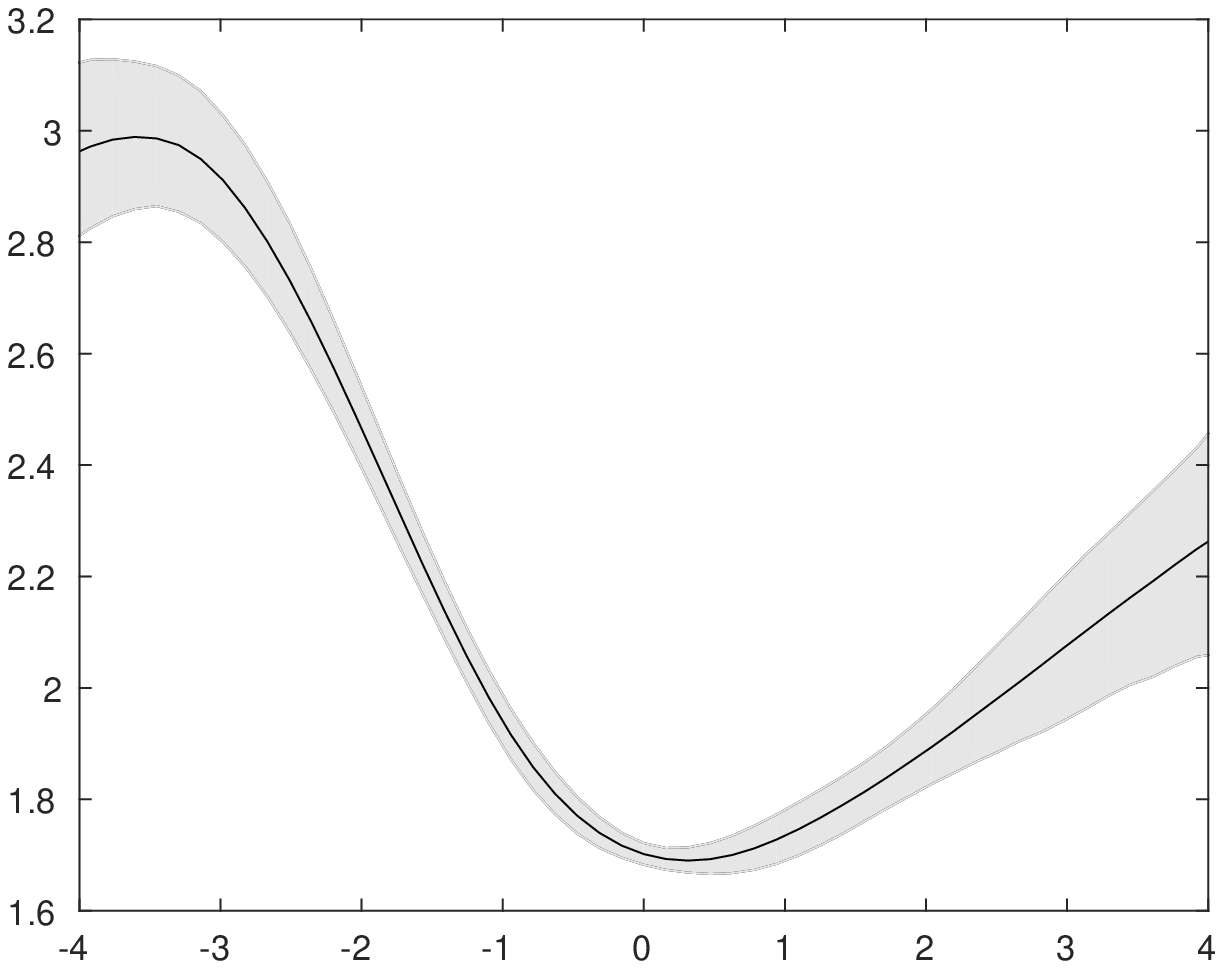}\\
\multicolumn{2}{c}{\multirow{1}{*}{STI}} \\
\\
\includegraphics[width=0.2\linewidth]{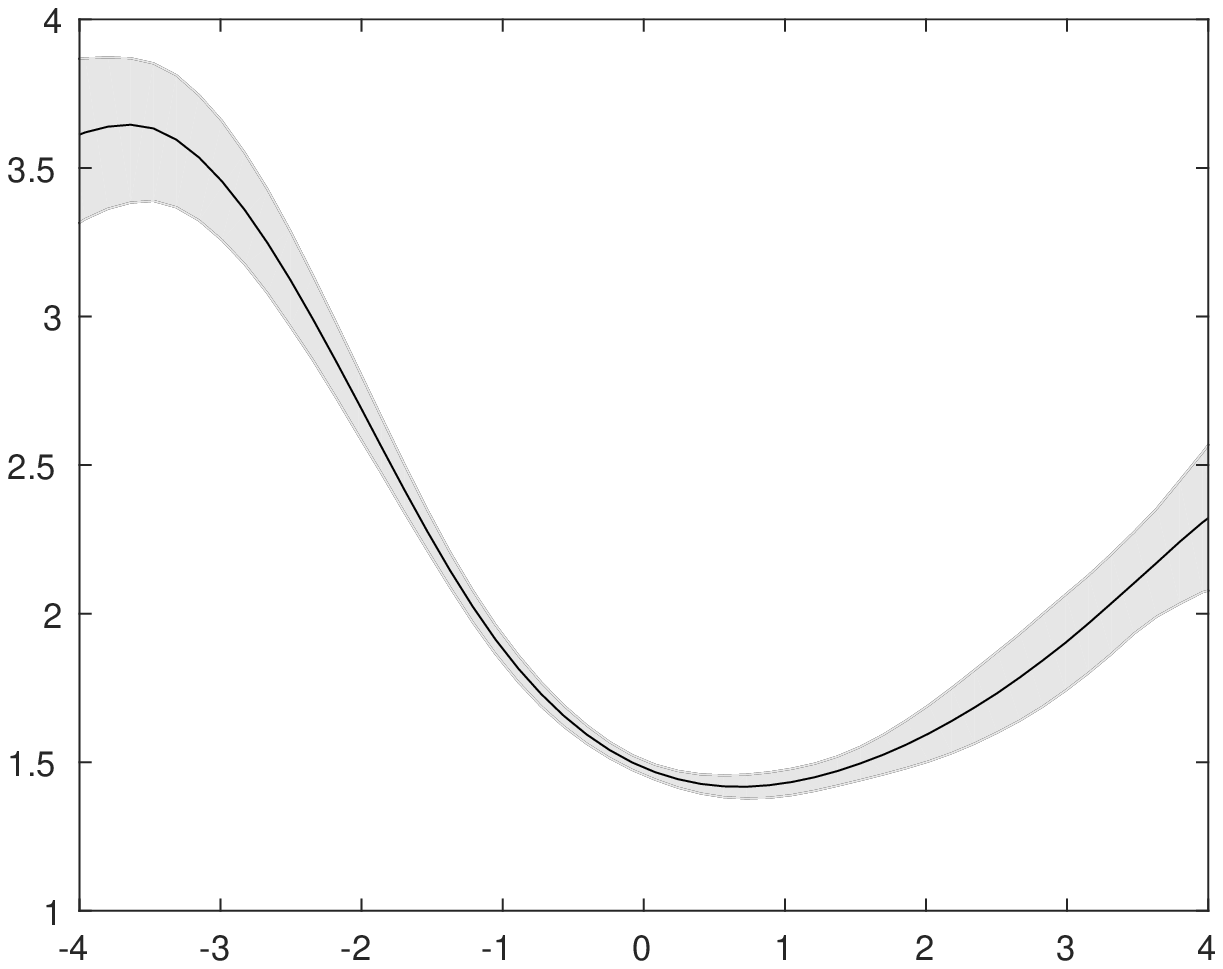}& \qquad
\includegraphics[width=0.2\linewidth]{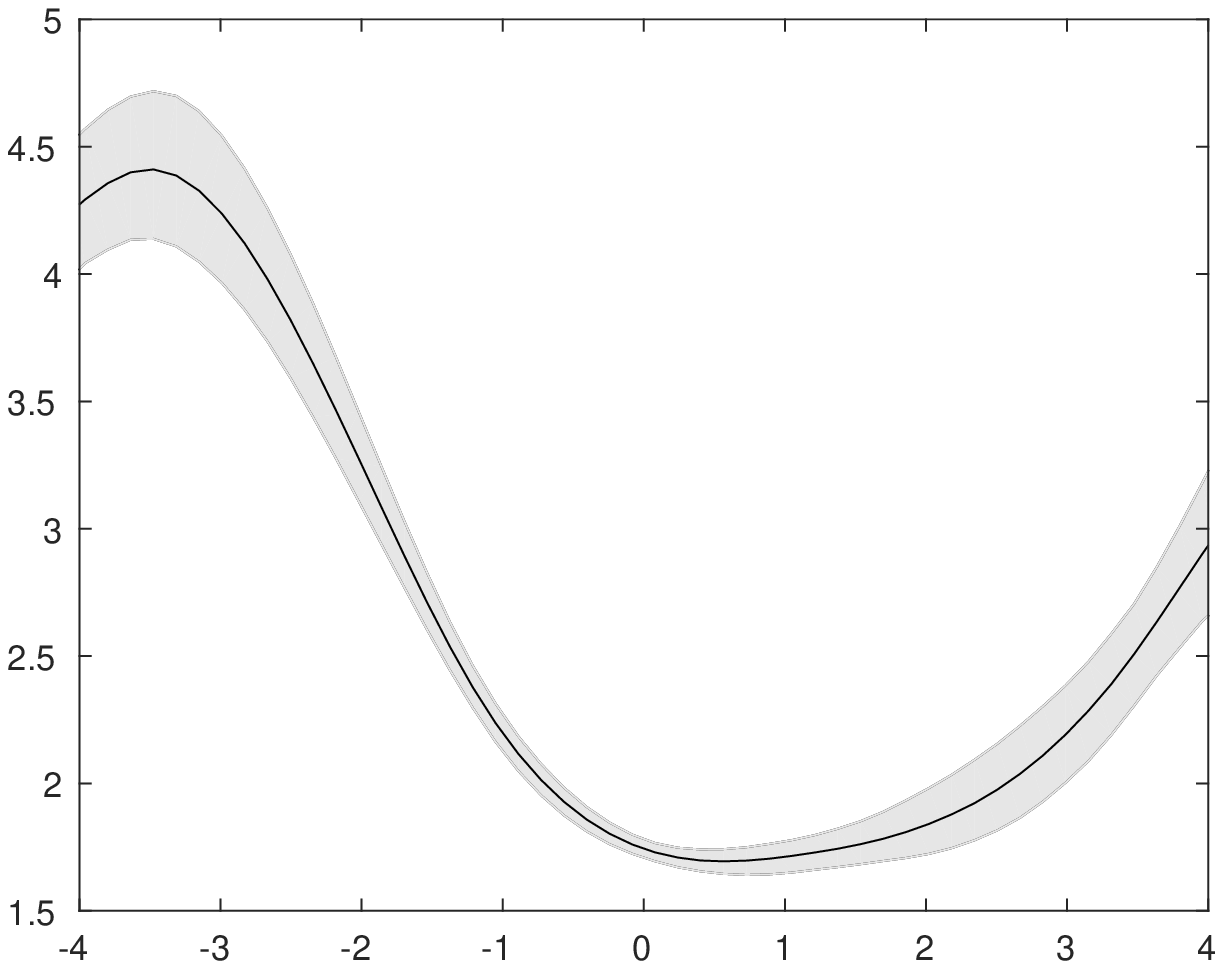}\\
\end{tabular}
\caption{Posterior estimate of the BNL-CARE NIC from the empirical application of section \ref{sec:empirical_applications}. Left panels exhibit the NIC (black line) along with the $95\%$ HPD regions (grey areas) at the quantile confidence levels $\tau=\left(0.05\right)$. Right panels exhibit the NIC (black line) along with the $95\%$ HPD regions (grey areas) at the quantile confidence levels $\tau=\left(0.01\right)$.}
\label{fig:NLCARE_NIC}
\end{center}
\end{figure}
%

\begin{figure}[ht!]
\begin{center}
\begin{tabular}{c}
\multicolumn{1}{c}{\multirow{1}{*}{Nasdaq $\tau=0.05$}} \\
\\
\includegraphics[width=0.9\linewidth]{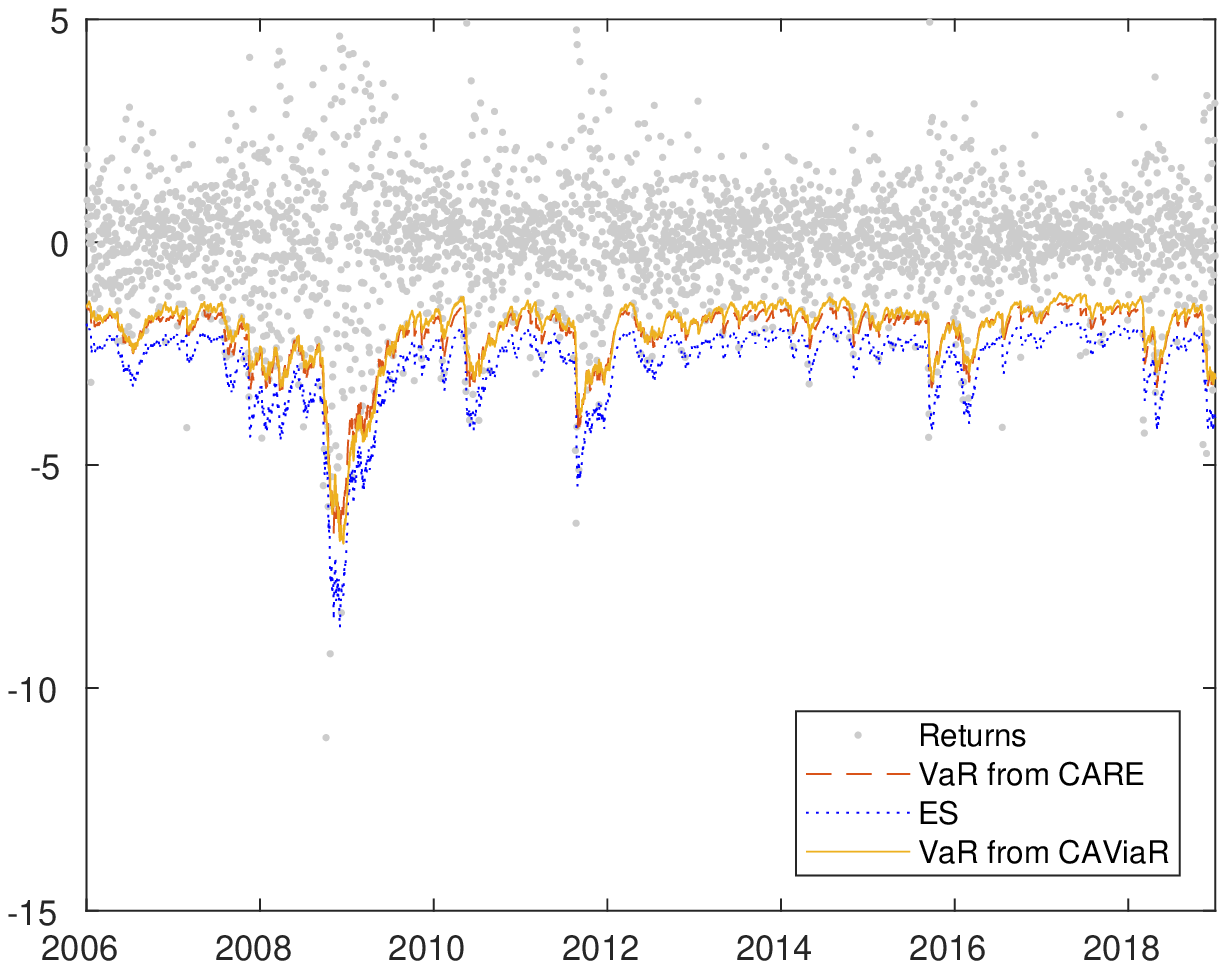}
\\
\multicolumn{1}{c}{\multirow{1}{*}{Nasdaq $\tau=0.01$}} \\
\\
\includegraphics[width=0.9\linewidth]{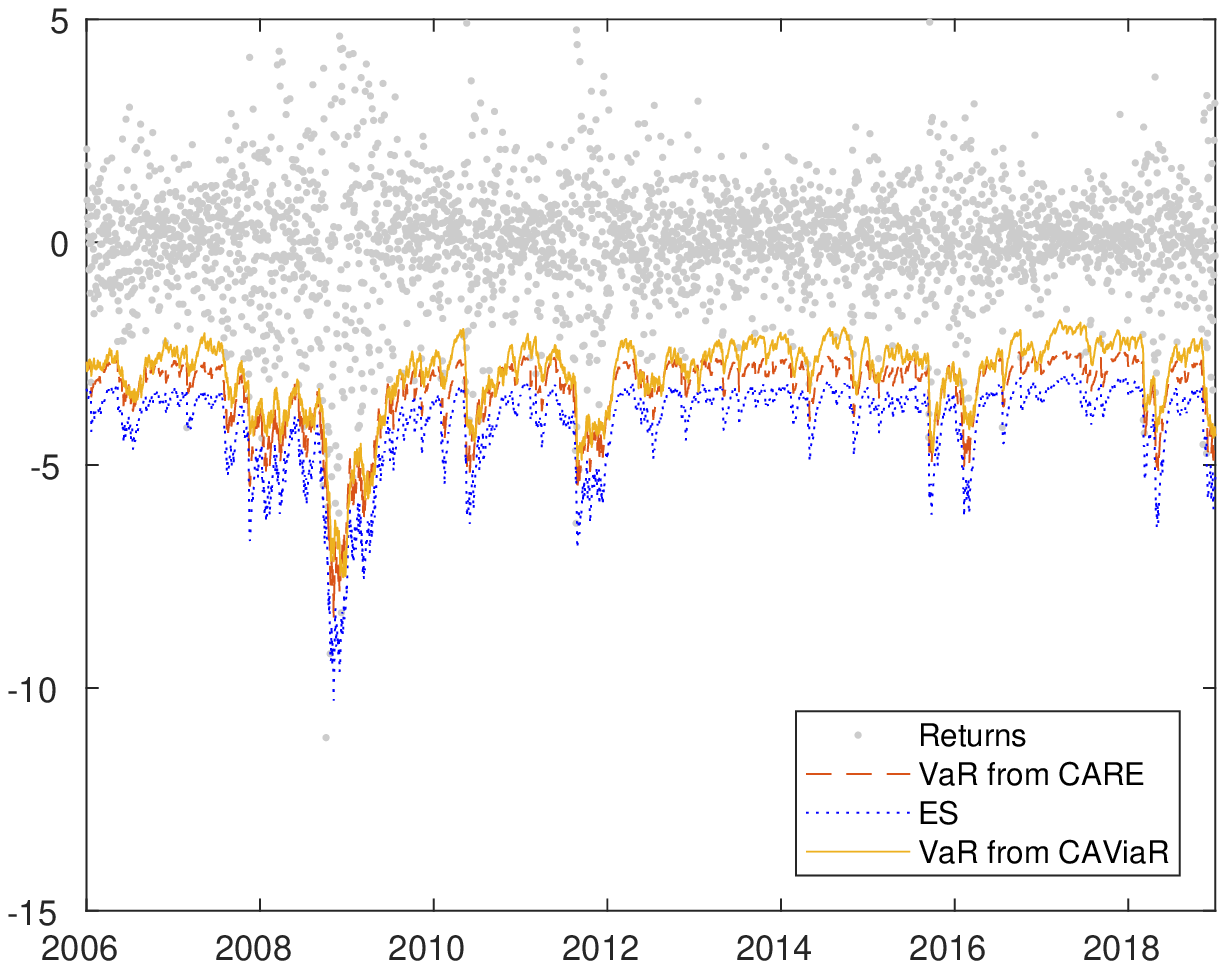}
\\
\end{tabular}
\caption{Posterior estimate of VaR and ES for the Nasdaq series, from the empirical application of section \ref{sec:empirical_applications}.}
\label{fig:NLCARE_NIC}
\end{center}
\end{figure}
%

\begin{figure}[ht!]
\begin{center}
\begin{tabular}{c}
\multicolumn{1}{c}{\multirow{1}{*}{KOREA SE $\tau=0.05$}} \\
\\
\includegraphics[width=0.9\linewidth]{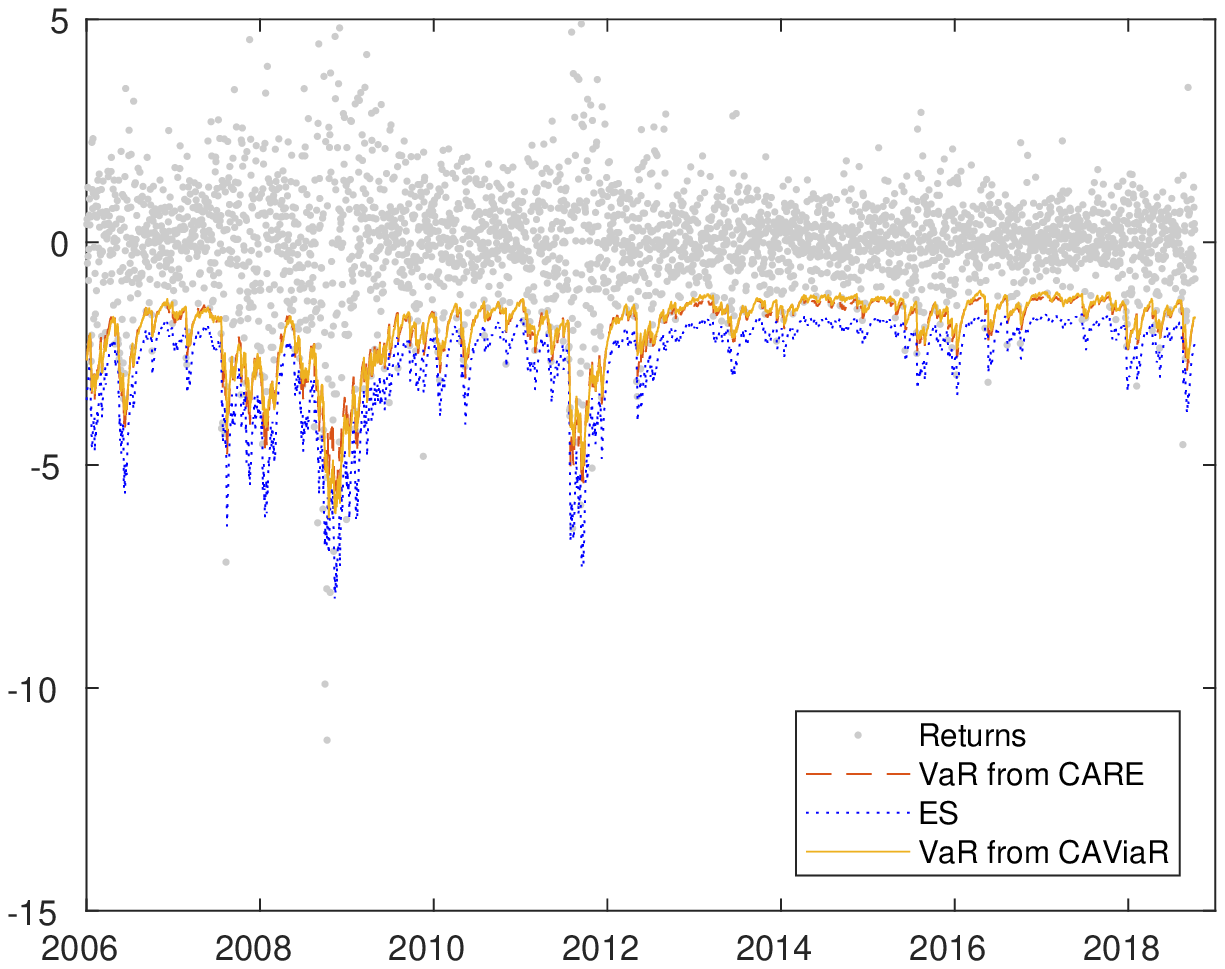}
\\
\multicolumn{1}{c}{\multirow{1}{*}{KOREA SE $\tau=0.01$}} \\
\\
\includegraphics[width=0.9\linewidth]{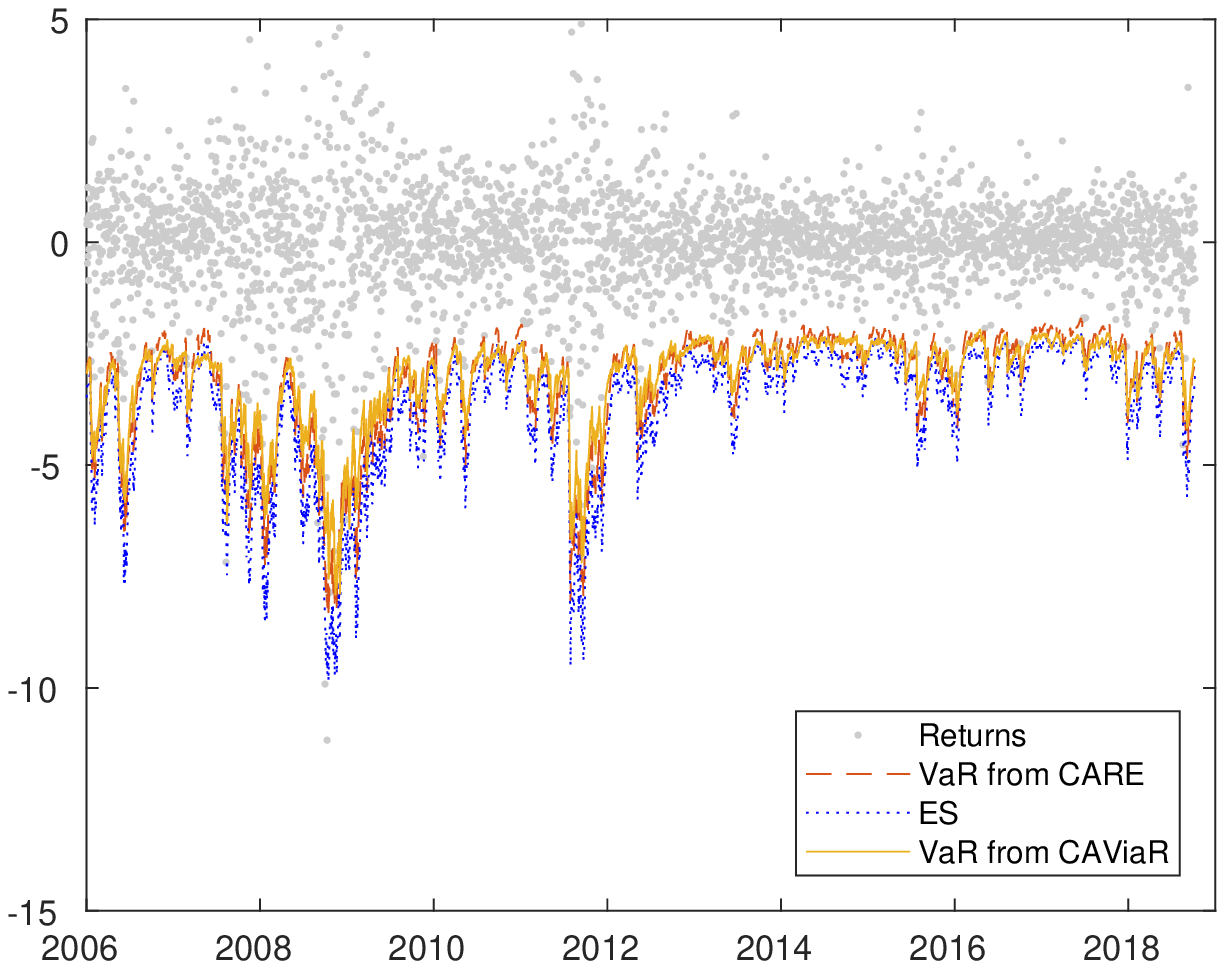}
\\
\end{tabular}
\caption{Posterior estimate of VaR and ES for the KOREA SE series, from the empirical application of section \ref{sec:empirical_applications}.}
\label{fig:NLCARE_NIC}
\end{center}
\end{figure}
%

\clearpage
\newpage
\bibliographystyle{apalike}
\bibliography{bnl_CARM_biblio}

\end{document}